\documentclass[aps,pra,groupedaddress,amsmath,amssymb,showpacs,twocolumn,floatfix]{revtex4-1}
\usepackage{graphicx}
\usepackage{bm,bbm,braket}
\usepackage{color}
\usepackage{times,txfonts}
\usepackage{hyperref}

\usepackage[caption=false]{subfig}

\newcommand{\brho}{\bm \rho}

\newcommand{\z}{\bm z}

\newcommand{\bpsi}{\bm \psi}

\def\Tr{\mathrm{Tr}}

\usepackage{booktabs,multirow,array}

\begin{document}

\title{A geometric approach to entanglement quantification with polynomial measures}

\author{Bartosz Regula}
\affiliation{$\mbox{School of Mathematical Sciences, The University of Nottingham, University Park,
    Nottingham NG7 2RD, United Kingdom}$}

\author{Gerardo Adesso}
\affiliation{$\mbox{School of Mathematical Sciences, The University of Nottingham, University Park,
    Nottingham NG7 2RD, United Kingdom}$}

\date{\today}

\begin{abstract}
  We show that the quantification of entanglement of any rank-2 state with any polynomial entanglement measure can be recast as a geometric problem on the corresponding Bloch sphere. This approach provides novel insight into the properties of entanglement and allows us to relate different polynomial measures to each other, simplifying their quantification. In particular, unveiling and exploiting the geometric structure of the concurrence for two qubits, we show that the convex roof of any polynomial measure of entanglement can be quantified exactly for all rank-2 states of an arbitrary number of qubits which have only one or two unentangled states in their range. We give explicit examples by quantifying the three-tangle exactly for several representative classes of  three-qubit states. We further show how our methods can be used to obtain analytical results for entanglement of more complex states if one can exploit symmetries in their geometric representation.
\end{abstract}


\maketitle

\section{Introduction}

Ever since entanglement was recognized as a useful resource in many quantum information protocols, there has been a consistent effort to develop a comprehensive framework for entanglement quantification \cite{plenio_2007,horodecki_2009}. However, the promising results in quantifying bipartite entanglement did not easily generalize to systems of more parties, where even for the simplest instance of three qubits only a handful of analytical results have been obtained in special cases \cite{eltschka_2014-1}.
In particular, the complex optimization problems involved in the quantification of any measure of multipartite entanglement are a major obstacle to obtaining a full understanding of the properties of entanglement in general, and to fulfilling a complete classification of useful mixed states based on their degrees of entanglement.

There are many inequivalent approaches to entanglement quantification, but one can nevertheless establish a set of basic rules that a function has to satisfy in order to quantify the resource character of entanglement. By an \emph{entanglement measure} we generally understand a function $E$ which vanishes only for separable states, is invariant under local unitary transformations, and is non-increasing under local operations and classical communication (LOCC) \cite{vedral_1997,vedral_1998,vidal_2000,horodecki_2009,eltschka_2014-1}, with the latter property meaning that $E$ is a so-called \emph{entanglement monotone}. Another often-imposed requirement is for $E$ to be a convex function \cite{vidal_2000}, which is satisfied by almost all measures of entanglement \cite{plenio_2007,eltschka_2014-1}, but is not necessary for LOCC monotonicity \cite{plenio_2005}.

Given these requirements, one can distinguish in particular two main classes of entanglement measures: the \emph{distance}-based measures, quantifying the distance of a given state $\rho$ to the set of all separable states according to some suitable (quasi)distance function \cite{vedral_1997,bengtsson_2007}, and the measures based on the \emph{convex roof}, where an entanglement monotone $E(\ket\psi)$ defined first on the set of pure states is then extended to the set of all mixed states by minimising its average value over all possible convex decompositions of the given state $\rho$ into pure states \cite{bennett_1996,uhlmann_1998}:
\begin{equation}
  \label{eq:convexroof}
  E(\rho) = \min_{\{p_i, \ket{\psi_i}\}} \sum_i p_i E(\ket{\psi_i}),
\end{equation}
for every decomposition $\rho = \sum_i p_i \ket{\psi_i}\bra{\psi_i}$ with $p_i \geq 0, \sum_i p_i = 1$. The decomposition(s) $\{p_i, \ket{\psi_i}\}$ realising the minimum in Eq.~(\ref{eq:convexroof}) is (are) called \emph{optimal}, and being able to find any such a decomposition in closed form for any given mixed state $\rho$ yields a full analytical quantification of the entanglement of $\rho$ according to the measure $E$. An important property of the convex roof is that a function obtained this way from a pure-state measure will always be an entanglement monotone on all states \cite{vidal_2000}. More intuitively, the convex roof procedure can be understood as extending the pure-state measure $E(\ket\psi)$ to mixed states ``as linearly as possible'' \cite{uhlmann_2010}, with $E(\rho)$ being in fact the largest convex function on the set of all mixed states which corresponds to $E(\ket\psi)$ on the set of pure states \cite{uhlmann_1998}. Although their purpose is to measure the same resource ---  entanglement --- the distance- and convex roof-based approaches are generally inequivalent and not directly related to each other. A notable link has been established for the so-called geometric measure of entanglement \cite{wei_2003}, whose convex roof extension was found to be equivalent to a distance-based measure based on fidelity \cite{streltsov_2010}.

An important class of pure-state entanglement measures is constituted by the \emph{polynomial measures}, based on homogeneous polynomial functions in the coefficients of a pure state $\ket{\psi}$ which are invariant under stochastic LOCC (SLOCC) \cite{eltschka_2014-1}. Any such a polynomial invariant $P$ of homogeneous degree $d$ can be written as
\begin{equation}
  P_d\,(c\, L \ket{\psi}) = c^d\, P_d\,(\ket{\psi}),
\end{equation}
for a constant $c >0$ and an invertible linear operator $L\in \text{SL}(m,\mathbb{C})^{\otimes n}$ representing a SLOCC transformation \cite{dur_2000} on each of a set of  $n$ $m$-dimensional systems. Then, one can take an appropriate power $p$ of the absolute value of any polynomial invariant $P$ to construct an entanglement measure on pure states,
\begin{equation}\label{eq:Edp}
  E_d^p(\ket{\psi}) = |P_d(\ket{\psi})|^p.
\end{equation}
For $n$-qubit states ($m=2$), the above expression defines a valid entanglement monotone provided  $d p \leq 4$  \cite{verstraete_2003, eltschka_2012}. Let us stress that, in the following, by $d$ we will always refer to the degree of the polynomial invariant $P_d$ itself, not the final homogeneous degree $d p$ of the measure obtained from it. The concept of polynomial invariants can be used to obtain entanglement measures for different types of entanglement in any number of qubits \cite{osterloh_2005,djokovic_2009} and qudits \cite{gour_2013}. Two particularly common monotones obtained in this way are the {\em concurrence} for two qubits \cite{hill_1997,wootters_1998} and the {\em three-tangle} for three qubits \cite{coffman_2000}. Notably, the convex roof extension of the concurrence can be quantified exactly for any system of two qubits \cite{wootters_1998}, although analytical solutions for the convex roof of the three-tangle have been found only in very special cases \cite{lohmayer_2006,eltschka_2008,jung_2009,jung_2009-1,he_2011,siewert_2012,viehmann_2012,eltschka_2012-1,regula_2016} and the first insights into the solutions of the even more complicated case of the convex roof of four-qubit polynomial measures have been obtained only recently \cite{jung_2015}.

In this paper, we develop a {\it geometric} approach to understanding and quantifying convex roof-extended polynomial measures of entanglement, establishing a link between geometric and algebraic methods for entanglement quantification. Our approach reveals common relations between different polynomial measures on pure states and allows for a simplification of the problem of evaluating their convex roof on mixed states. Specifically, for all rank-2 states of multipartite systems, whose range can be represented geometrically as a Bloch sphere, we explicitly demonstrate that computing the entanglement (according to any such polynomial measure) of arbitrary pure states on the surface of the sphere corresponds to calculating a product of Euclidean distances from a finite set of unentangled states (roots of the polynomial measure) on the sphere. This allows us to look at the problem of quantifying entanglement and evaluating the convex roof for any mixed state inside the sphere differently --- by employing only elementary Euclidean geometry. When the considered polynomial admits no more than two distinct roots (with equal multiplicities), we solve the problem completely for all the states in the corresponding Bloch ball. In particular, we investigate the geometric structure of the two-qubit concurrence, and show that in relevant cases the same structure is shared by polynomial measures of higher degrees, such as the three-tangle for three qubits. This allows us to evaluate the convex roof of the three-tangle exactly in a variety of rank-2 states of three qubits (for which no solution was available so far, to our knowledge). We describe such instances in detail and also show possible ways to extend our geometric method to more general states.

The paper is organized as follows. In section~\ref{sec:geometry} we describe the geometry of polynomial entanglement measures for arbitrary rank-2 states and prove its relation to products of distances. In section~\ref{sec:concurrence} we reassess the convex roof problem for the concurrence of two-qubit states in purely geometric terms. In section~\ref{sec:tangle} we apply the same methods to provide an exact solution to the convex roof problem for the three-tangle of all rank-2 mixed states of three qubits with no more than two unentangled pure states (two roots) in their range; we further provide a complete classification of such states. In section~\ref{sec:general} we discuss extensions to more general polynomial measures of entanglement. We summarize our results in section~\ref{sec:concl}.

\begin{figure*}[t]
  \centering
  \includegraphics[width=13cm]{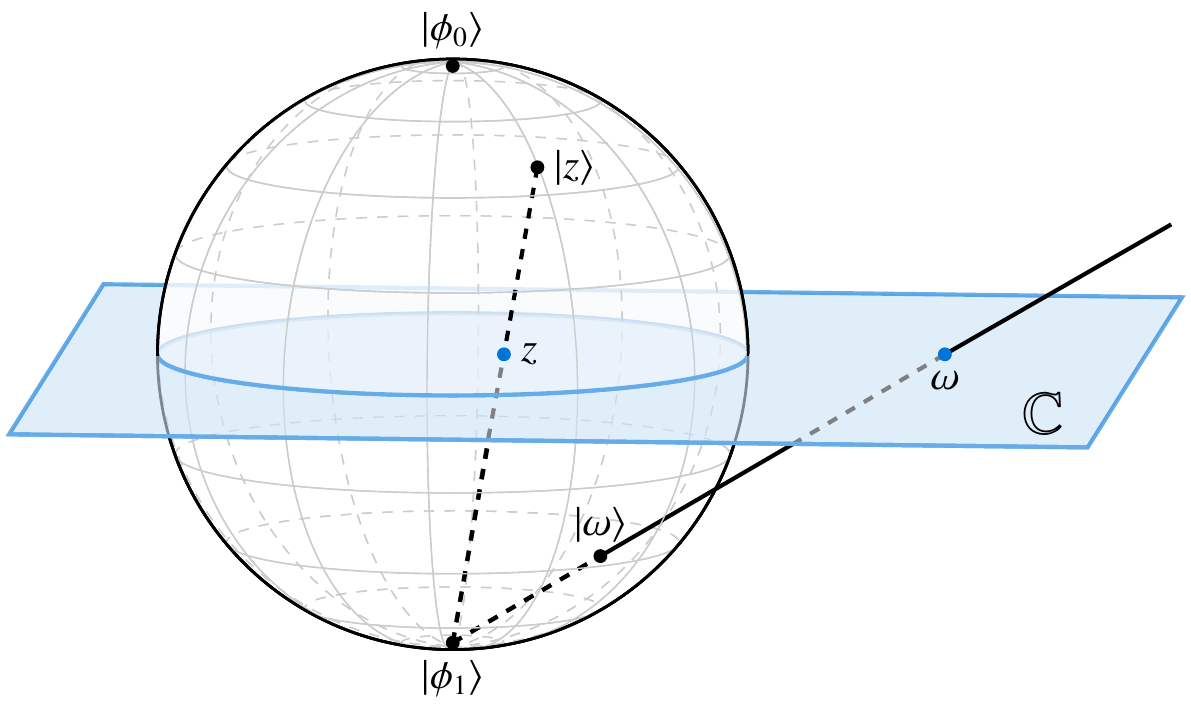}
  \caption{The Bloch sphere as the stereographic projection onto the complex plane (shaded blue plane). Every pure state $\ket{\omega}$ with a Bloch vector on the sphere corresponds to the point $\omega \in \mathbb{C}$ which lies at the intersection of the plane with the line containing the Bloch vectors of $\ket{\omega}$ and the projection point $\ket{\phi_1}$.  States in the upper hemisphere are mapped to the region of the plane inside the sphere, and states in the lower hemisphere are mapped outside of it. The projection point itself, corresponding to the state $\ket{\phi_1}$, is mapped to a point at infinity.}
  \label{fig:projection}
\end{figure*}

\section{Geometric setting}
\label{sec:geometry}

In this paper, we will focus our attention on rank-2 states of an arbitrary number of qudits.
Given a rank-2 quantum state $\rho$, it can always be written in its spectral decomposition into the two eigenvectors $\{\ket{\phi_0}, \ket{\phi_1}\}$ corresponding to the two non-zero eigenvalues $\{\lambda_0, \lambda_1\}$,
\begin{equation}
  \rho = \lambda_0 \ket{\phi_0}\bra{\phi_0} + \lambda_1 \ket{\phi_1}\bra{\phi_1},
\end{equation}
with $\lambda_0+\lambda_1=1$.
We can then visualize the range of $\rho$ as the Bloch sphere with the two eigenvectors as poles, and any state $\ket{\omega}$ on the surface of the sphere as corresponding (up to normalization) to a linear combination of the two eigenvectors, $\ket{\omega}=\ket{\phi_0} + \omega \ket{\phi_1}$ for some $\omega \in \mathbb{C}$, with $\ket{\phi_1}$ itself corresponding to a point $\omega$ at infinity. This, in fact, can be understood as the stereographic projection from the Bloch sphere onto the extended complex plane $\hat{\mathbb{C}} = \mathbb{C} \cup \{\infty\}$ (Fig.~\ref{fig:projection}).

Let $E_d$ denote a polynomial entanglement measure based on a polynomial invariant $P_d$ of degree $d$, as in Eq.~(\ref{eq:Edp}). Let $\{\ket{\phi_0}$, $\ket{\phi_1}\}$ be the two eigenvectors of a rank-2 state $\rho$ corresponding to its non-zero eigenvalues.
Then, the entanglement $E_d$ of any state $\ket{\omega}$ in the range of $\rho$ can  be expressed as the absolute value of a polynomial in one  complex  variable $\omega$:
\begin{equation}\label{eq:expression}
  E_d\,\Big( \ket{\phi_0} + \omega \ket{\phi_1} \Big) = N \prod_{i=1}^{d} |\omega - z_i|,
\end{equation}
where $N$ is a normalization constant, and $\{z_1, \ldots, z_d\}$ are the roots of the polynomial $E_d$ in $\hat{\mathbb{C}}$, defined by
\begin{equation}
  E_d\,\Big(\ket{\phi_0} + z_i \ket{\phi_1}\Big) = 0.
\end{equation}
The convex hull of the points on the Bloch sphere corresponding to these coefficients $\{z_i\}$  defines the \emph{zero polytope} \cite{lohmayer_2006,osterloh_2008}. This concept provides a useful representation of the set of all separable states within the considered Bloch sphere,  
since the entanglement as measured by the convex roof extension of $E_d$ has to vanish for any convex combination of pure separable states (i.e., for any mixed state inside the zero polytope), but will not vanish outside of their convex hull.

We proceed by explicitly normalising the state $\ket{\omega} = \ket{\phi_0} + \omega \ket{\phi_1}$ and additionally dividing the expression (\ref{eq:expression}) by the normalization factors of all the states $\{\ket{z_i}\}$, to obtain
\begin{equation}\label{eq:normalised}
  \frac{1}{\prod_i \sqrt{1 + |z_i|^2}} E_d\left( \frac{\ket{\phi_0} + \omega \ket{\phi_1}}{\sqrt{1+|\omega|^2}} \right) = \frac{N \prod_{i} |\omega - z_i|}{\sqrt{1+|\omega|^2}^{\,d} \prod_i \sqrt{1 + |z_i|^2}}.
\end{equation}
Let us now recall the definition of the chordal distance in the stereographic projection, that is, the Euclidean distance between points $\bm x_k$ on the sphere $S^2$ corresponding to their stereographic projection $\xi_k$ onto the extended complex plane $\hat{\mathbb{C}}$ \cite{caratheodory_1954} (see Fig.~\ref{fig:projection}):
\begin{equation}
  \| \bm x_1 - \bm x_2 \| = \frac{2\,|\xi_1 - \xi_2|}{\sqrt{1 + |\xi_1|^2}\sqrt{1 + |\xi_2|^2}},
\end{equation}
with
\begin{equation}
  \| \bm x_1 - \bm \infty \| = \frac{2}{\sqrt{1 + |\xi_1|^2}}.
\end{equation}
Since the Bloch sphere is precisely an (inverse) stereographic projection of $\hat{\mathbb{C}}$ onto a sphere $S^2$, we identify the right-hand side of Eq.~(\ref{eq:normalised}) with a product of distances between pure-state Bloch vectors defined by their respective complex coefficients. This leads to the simple relation
\begin{equation}
  \begin{aligned}
    E_d\left( \frac{\ket{\phi_0} + \omega \ket{\phi_1}}{\sqrt{1+|\omega|^2}} \right) &= \frac{N}{2^d} \prod_i \sqrt{1+|z_i|^2} \prod_i \| \bm{\omega} - \bm{z}_i \|\\
    &= N_\rho \prod_i \| \bm{\omega} - \bm{z}_i \|
  \end{aligned}
\end{equation}
where $\bm{\omega}$ and $\bm{z}_i$ are the Bloch vectors corresponding to $\omega$ and each $z_i$, respectively, and we have introduced the normalization constant $N_\rho$ explicitly dependent on the properties of the density matrix of the state $\rho$ whose eigenbasis defines the Bloch sphere.

We have thus shown that the entanglement of any pure state $\ket{\psi}$ in the range of a rank-2 state $\rho$ can be quantified with any polynomial measure $E_d$ considering only the product of Euclidean distances between the Bloch vector of the state $\ket{\psi}$ and the Bloch vectors of the $d$ roots of the measure $E_d$ on the corresponding Bloch sphere. The value of the constant $N_\rho$ rescales the entanglement in the range of a given state $\rho$ appropriately, and can be obtained by an explicit evaluation of $E_d$ on any pure state on the Bloch sphere with non-zero entanglement, which is of course straightforward because $E_d$ is defined as a given polynomial function in the coefficients of any pure-state vector.

This relation lets us immediately conclude that all polynomial measures of entanglement for rank-2 states share a common geometric structure on the surface of the Bloch sphere, and since the method of extending the measures to the inside of the Bloch sphere by the convex roof only depends on the values on the surface, the geometry will also remain equivalent on the inside. In particular, given any two measures based on a polynomial of the same degree and admitting the same set of roots, their geometric structure in the Bloch ball will be the same, and their values for all states will be equal up to the normalization constant. The geometric structure is, in fact, preserved between measures of a different degree in degenerate cases --- for example, as we will show in detail later, a measure of a higher degree can be reduced to a measure of lower degree if some of its polynomial roots $\{z_i\}$ are repeated.

We further note that the particular case of the Bloch ball in $\mathbb{R}^3$ means that many measures of distances commonly employed in quantum information, such as the trace distance or the Hilbert-Schmidt distance, are actually equivalent to the Euclidean distance up to a constant factor \cite{bengtsson_2007} --- a fact that no longer applies in higher dimensions, but is nevertheless useful in the present case of rank-2 states.
Additionally, we remark that the product of distances on the sphere is a rather well-studied problem in mathematics, although usually only in terms of maximising or minimising the product with regards to the arrangement of points on the surface \cite{whyte_1952,wagner_1989,rakhmanov_1994}; in our setting the points on the surface are instead fixed, being determined by the roots of $E_d$.

Let us now consider explicit applications of this result. We will in particular re-derive known findings for the concurrence of two-qubit states from a purely geometric perspective, and use the same methods to obtain new results for the three-tangle of rank-2 mixed states of three qubits.

\begin{figure*}[t]
  \centering
  \subfloat[]{\includegraphics[width=8cm]{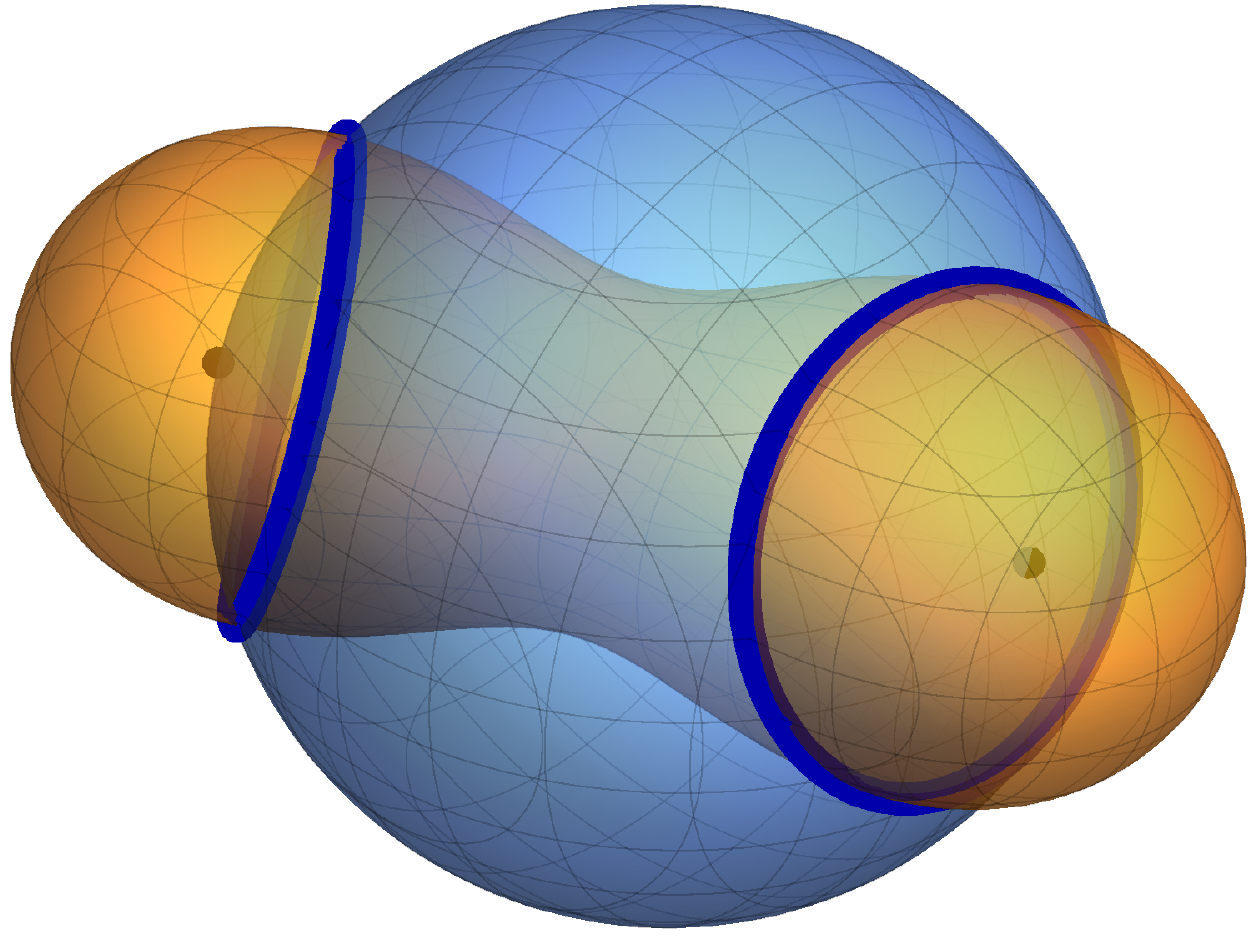}}
  \subfloat[]{\includegraphics[width=8cm]{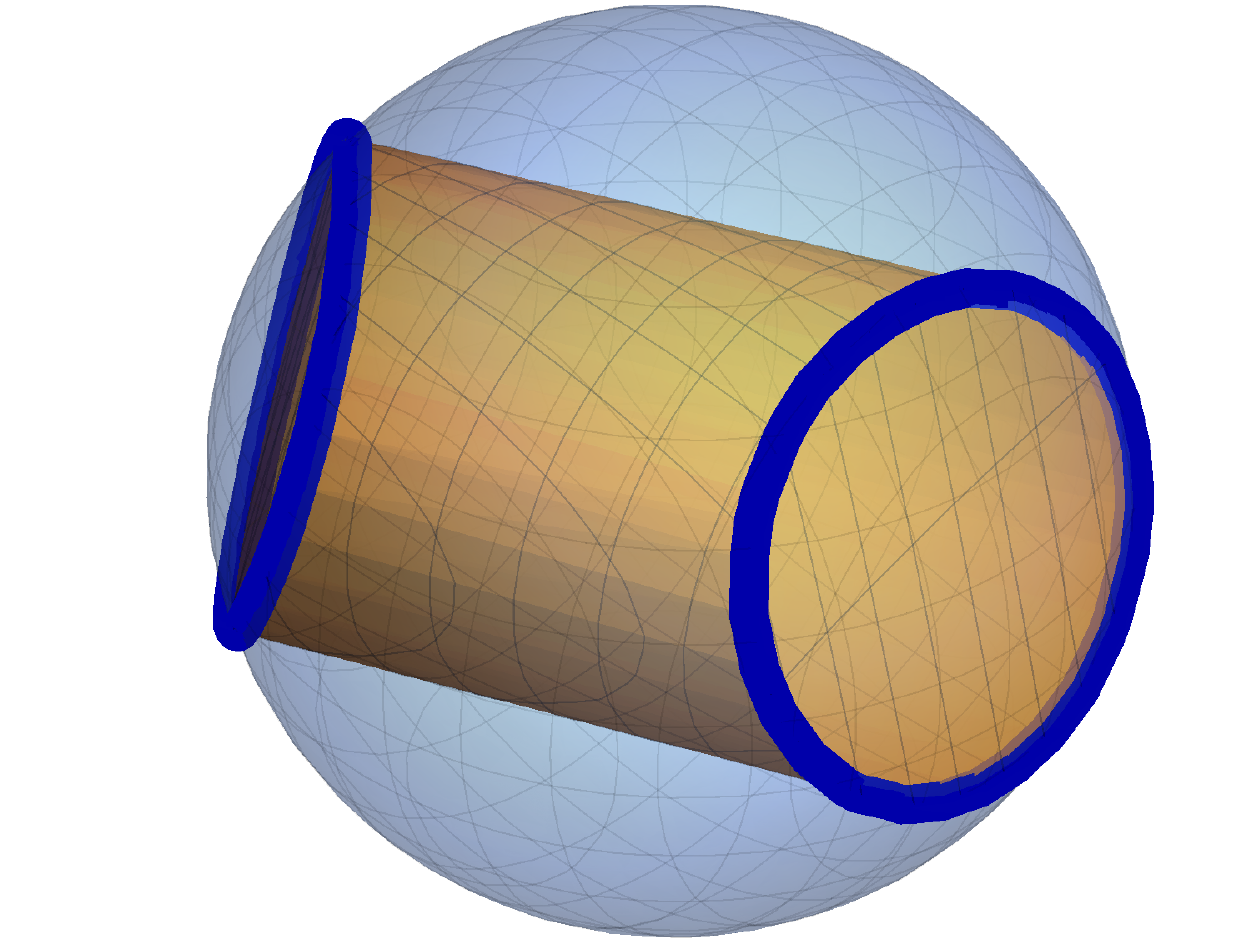}}
  \caption{(a) The peanut-shaped surface is the locus of points with a constant product of distances to the two foci $\bm z_1, \bm z_2$. The intersection of the surface with the unit sphere is then a curve of constant concurrence, here plotted for $C = N_\rho$ and a choice of the root points.\newline\hspace{\textwidth}(b) The convex hull of the curves of constant concurrence. Each mixed state in the convex hull admits a flat decomposition into pure states with the same value of concurrence.}
  \label{fig:concurrence}
\end{figure*}

\section{Concurrence of two qubits}
\label{sec:concurrence}

The concurrence $C$ is defined for a pure state $\ket{\psi} \in \mathbb{C}^2 \otimes \mathbb{C}^2$ of two qubits as \cite{hill_1997,wootters_1998,wootters_2001}
\begin{equation}
  C\big(\ket\psi\big) = 2\, \big| \psi_{00} \psi_{11} - \psi_{01} \psi_{10} \big|
\end{equation}
where $\psi_{ij}$ are the coefficients of the state $\ket\psi$ in the computational basis. Since it is explicitly a polynomial entanglement measure of degree 2, we can follow the results of section \ref{sec:geometry} to obtain the concurrence of any pure state in the range of a rank-2 state $\rho$ by
\begin{equation}
  C\Big(\ket{\phi_0} + \omega \ket{\phi_1}\Big) = N_\rho\, \| \bm{\omega} - \bm{z}_1 \|\,  \|\bm{\omega} - \bm{z}_2\|
\end{equation}
for the two polynomial roots $z_1, z_2$.

To investigate the structure of this function for all mixed states, we would first like to find curves of constant concurrence on the surface of the Bloch sphere. We thus consider the locus of points such that their product of distances from the two roots is constant. Such curves defined in the plane $\mathbb{R}^2$ are called Cassini curves or Cassini ovals \cite{cassini_1684,needham_1998}, and we visualize their extension as surfaces in $\mathbb{R}^3$ in Fig.~\ref{fig:concurrence}(a). The curves of constant concurrence are then the intersections of these surfaces with the Bloch sphere.

The convex combinations of the two root points form the zero polytope inside the sphere, which is reduced to a \emph{zero line} joining $\bm z_1$ and $\bm z_2$ and forming the axis of the Cassini surface. We note that the concurrence on the surface of the sphere is symmetric around the plane perpendicular to the zero line and containing the midpoint between $\bm z_1$ and $\bm z_1$. Therefore, any mixed state with a Bloch vector $\bm \rho$ inside the sphere can be decomposed into two pure states of equal concurrence, lying on opposite sides of the sphere with regards to the plane of symmetry.

\begin{figure}[t]
  \centering
  \includegraphics[width=8cm]{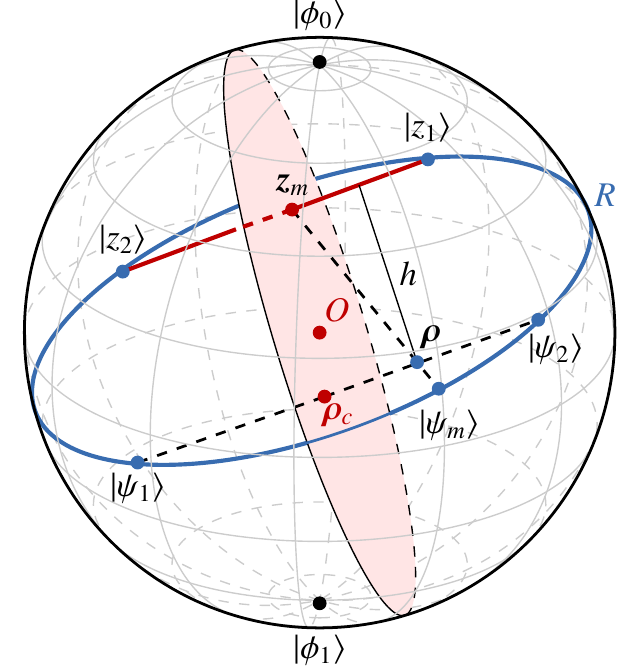}
  \caption{Quantifying the concurrence of any state $\rho$ inside the Bloch sphere of a two-root state. On the surface of the sphere, the two root states $\ket{z_1}$ and $\ket{z_2}$ are the only separable pure states. The line joining them then defines the zero line (thick blue line), and the perpendicular plane bisecting the zero line is the plane of symmetry of concurrence (shaded red plane). The state $\rho$ admits a decomposition into two pure states of equal concurrence, $\ket{\psi_1}$ and $\ket{\psi_2}$. The value of concurrence of $\rho$ can then be expressed in terms of the distance from the zero line ($h$) and the radius of the small circle of the sphere containing $\rho$ and the zero line ($R$). See main text for details.}
  \label{fig:plot}
\end{figure}

Let us now work in purely geometric terms and disregard momentarily the scaling constant $N_\rho$. Let $P$ denote the product of distances from a chosen point to the root points:
\begin{equation}
  P(\bm\rho\,|\, \bm z_1, \cdots, \bm z_n) = \prod_i \| \bm{\rho} - \bm{z}_i \|.
\end{equation}
In the case under study ($n=2$), we first consider as candidate convex roof a function $f(\rho)$ which corresponds to the product of distances $P(\bm \rho \,|\, \bm z_1 \bm z_2)$ for all points on the surface of the sphere, and assigns to each mixed state $\rho$ inside the sphere the value of $P$ for the pure states in its aforementioned convex decomposition along a line parallel to the zero line $\z_1 \z_2$ into two states $\ket{\psi_1}$ and $\ket{\psi_2}$ of equal $P$, as indicated in Fig.~\ref{fig:plot}. Using the law of sines and elementary geometry, we obtain
\begin{equation}
  f(\rho) = 2\, R\, h
\end{equation}
where $R$ is the radius of the small circle of the sphere in the plane containing $\bm\rho \bm z_1 \bm z_2$, and $h$ is the distance of $\bm \rho$ from the zero line $\bm z_1 \bm z_2$ (see Fig.~\ref{fig:plot}). $R$ can be obtained as $\sqrt{1 - s^2}$, where $s$ is the distance of the plane containing the small circle to the center of the sphere. By construction, $N_\rho f(\rho)$ corresponds to the function considered by Hill and Wootters and therefore to the mixed-state concurrence \cite{hill_1997}, but we will now prove this result explicitly in the geometric approach to justify its further generalizations.

Noting that $f$ is constant in the direction parallel to the zero line, we limit ourselves to the plane of symmetry of the sphere, and introduce Cartesian coordinates $(x,y)$ centered at the point of intersection of this plane with the zero line. Let $\rho_c = (x_\rho,y_\rho)$ denote the projection of any state $\rho$ along the constant direction onto the plane (see Fig.~\ref{fig:plot}). We then have
\begin{equation}
  \label{eq:ellipse}
  f(x_\rho, y_\rho) = \sqrt{(1-y_O^2) x_\rho^2 + (1-x_O^2) y_\rho^2},
\end{equation}
where $(x_O, y_O)$ are the coordinates of the origin of the sphere. It is now explicit that $f$ is in fact a norm in this plane, and is therefore convex \cite{rockafellar_1970}. Since $f(\rho)$ is constant in the direction perpendicular to the plane of symmetry, it follows that $f(\rho)$ is convex on the whole Bloch ball. Additionally, we note that Eq.~(\ref{eq:ellipse}) is the equation of an ellipse with the semi-major and semi-minor axes given by, respectively, $f(\rho)\left(1-y_O^2\right)^{-1/2}$ and $f(\rho)\left(1-x_O^2\right)^{-1/2}$.

Now, assume that there exists a function $f'(\rho)$, corresponding to a different decomposition of $\rho$, which also reduces to $P(\bm \rho \,|\, \bm z_1 \bm z_2)$ on the pure states and is also convex inside the Bloch sphere, but is such that $f'(\rho) > f(\rho)$ for at least one state $\rho$. Consider then the previously introduced decomposition of $\rho$ along the line parallel to the zero line as $\rho = p\ket{\psi_1}\bra{\psi_1} + (1-p)\ket{\psi_2}\bra{\psi_2}$. By convexity of $f'$ we then get
\begin{equation}
  \begin{aligned}
    f'(\rho) &\leq p f'(\ket{\psi_1}) + (1-p) f'(\ket{\psi_2})\\
    &= p f(\ket{\psi_1}) + (1-p) f(\ket{\psi_2}) \\
    &= f(\rho),
  \end{aligned}
\end{equation}
which is a contradiction. Hence $f$ is the largest convex function on the mixed states corresponding to $P$ on the surface of the sphere, and therefore constitutes the exact convex roof extension of the function to the whole ball \cite{uhlmann_1998}. Identifying the concurrence $C(\rho)$ with $N_\rho f(\rho)$, we obtain
\begin{equation}
  \label{eq:concurrence}
  C(\rho) = 2\, N_\rho \, R\, h
\end{equation}
as the convex roof extension of the concurrence to all rank-2 mixed states of two qubits. We note the explicit dependence of the entanglement of a state on its distance from the set of separable states in the Bloch sphere, establishing a link between geometric and algebraic (polynomial) approaches to entanglement quantification as advocated in this paper. 

We can now obtain an easily computable formula  for the concurrence of any rank-2 state $\rho$ in the Bloch sphere by considering a different convex decomposition. Choosing any point $\z_m$ lying on the zero line (including the roots $\z_1, \z_2$ themselves), we take the decomposition of $\rho$ into the (possibly mixed) separable state with the Bloch vector at $\z_m$ and the pure state $\ket{\psi_m}$ whose Bloch vector $\bm \psi_m$ lies on the line containing $\brho$ and $\z_m$ (see Fig.~\ref{fig:plot}). By elementary geometry, the value of concurrence in this decomposition is equal to Eq.~(\ref{eq:concurrence}), and so the concurrence of $\rho$ can be obtained as
\begin{equation}
  \label{eq:concurrencerdl}
  C(\rho) = C\big(\ket{\,\psi_m}\big)\, \frac{\| \brho - \z_m \|}{\| \bm \psi_m - \z_m \|},
\end{equation}
where the distances on the right-hand side can be evaluated using the trace or Hilbert-Schmidt distances, providing a computable formula in terms of the density matrices of the states. We remark that if we choose the point $\z_m$ to be defined as the projection of $\brho$ onto the zero line, the expression in Eq.~(\ref{eq:concurrencerdl}) corresponds to the so-called best separable approximation for the concurrence of $\rho$ \cite{lewenstein_1998}, showing that the bound given by this approximation is in fact tight on all rank-2 states. If, instead, we take $\z_m$ to be the midpoint between $\z_1$ and $\z_2$ (as in Fig.~\ref{fig:plot}), we get the unoptimized formula of the best zero-$E$ approximation \cite{rodriques_2014}, showing that this approximation (even without further optimization \cite{rodriques_2014, osterloh_2016}) is tight for all rank-2 states.

Because the optimal decompositions into two states with equal concurrence form straight lines through the sphere and we have shown their cross-sections to form ellipses, the surfaces of constant concurrence \emph{inside} the Bloch ball are elliptic cylinders around the zero line, as first noted in \cite{hill_1997}. These can be understood as the convex hull of the curves of constant concurrence. Explicitly, we consider such a curve on the surface of the sphere and extend the corresponding function by straight lines through the inside of the ball, as in Fig.~\ref{fig:concurrence}(b). This is then a surface of constant concurrence, because each mixed state on this surface can be decomposed into two pure states with equal concurrence. In fact, the same geometric structure and optimal decompositions are shared by {\it all} polynomial measures for states with only one or two distinct polynomial roots (with equal multiplicities), i.e., for states whose zero polytope reduces to a zero line as in the case of the concurrence for all rank-2 states of two qubits. Analogous constructions can also be obtained for general polynomial measures of a higher degree, although such decompositions are not always optimal, as we will explicitly show later.

Also worth noting is the fact that the geometric expressions for concurrence in Eqs.~(\ref{eq:concurrence}) and (\ref{eq:concurrencerdl}) can simplify to more straightforwardly computable formulas in several cases. If we have a state $\rho_1$ with only one root $z = z_1 = z_2$, the concurrence in the state's Bloch ball reduces to $C(\rho_1) = 2\, N_{\rho_1} \, h_c$ where $h_c$ is the distance from the Bloch point $\bm \rho_1$ to the plane tangent to the sphere at $\bm z$, or alternatively the distance from $\bm z$ to the center of the plane containing $\bm \rho_1$ and perpendicular to the axis going through $\bm z$ (see \cite{regula_2016}). This function is clearly linear through the Bloch ball in one direction and constant through the other two directions, meaning that the concurrence is an affine function throughout the Bloch sphere and the convex roof problem requires no minimization as it is constant for all convex decompositions, as has been shown more explicitly in \cite{regula_2016}. The state $\ket{z'}$ with a Bloch vector $\bm z'$ antipodal to $\bm z$ is then the maximally entangled state in the range of $\rho_1$. In terms of density matrix coefficients, this gives
\begin{equation}
  \begin{aligned}
    C(\rho_1) &= 2\, N_{\rho_1} \, \left| 1 - \braket{z \,|\, \rho_1 |\, z} + \braket{z' \,|\, \rho_1 |\, z'}\right|\\
    &= \frac{1}{2}\, C\big(\ket{z'}\big) \, \left| 1 - \braket{z \,|\, \rho_1 |\, z} + \braket{z' \,|\, \rho_1 |\, z'}\right|.
  \end{aligned}
\end{equation}

Another simplified expression occurs for states $\rho_d$ whose two root states $\ket{z_1}$ and $\ket{z_2}$ are orthogonal, $\langle z_1|z_2\rangle=0$,  meaning that the points $\z_1$ and $\z_2$ are antipodal to each other on the surface of the sphere. This means that the plane containing the zero line always contains the origin, and we have $R=1$. Any state $\ket{z''}$ whose Bloch vector lies on the great circle of the sphere perpendicular to the zero line then has the largest entanglement in this Bloch sphere. Hence
\begin{equation}
  \begin{aligned}
    C(\rho_d) &= 2 N_{\rho_d} \, h\\
    &= 2\, C\big(\ket{z''}\big)\,\left| \braket{z_1 \,|\, \rho_d \,|\, z_2}\right|.
  \end{aligned}
\end{equation}
In this case the entanglement reduces, up to normalization, to the $l_1$-norm of coherence \cite{baumgratz_2014} in the eigenbasis of the two roots $\{\ket{z_1}, \ket{z_2}\}$.

More generally, let us remark once more that the concurrence of any two-qubit mixed state is already computable in closed form thanks to Wootters' formula \cite{wootters_1998}, yet our geometric reformulation of the problem provides insights which will be particularly precious in the more complicated case of polynomial measures of entanglement for multipartite systems, as we show in the next section.

\section{Three-tangle of three qubits}
\label{sec:tangle}

The basic idea of the previous section can be readily extended to $\tau$, the polynomial measure of genuine tripartite entanglement for three qubits known as the residual tangle, or {\it three-tangle} \cite{coffman_2000}. For any pure state $\ket{\psi} \in \mathbb{C}^2 \otimes \mathbb{C}^2 \otimes \mathbb{C}^2$, we can write it as
\begin{equation}
	\begin{aligned}
  \tau\big(\ket\psi\big) = &4\Big|\, \psi_{000}^2\, \psi_{111}^2+\psi_{001}^2\, \psi_{110}^2+\psi_{010}^2\, \psi_{101}^2+\psi_{100}^2\, \psi_{011}^2\\
  &-2\,\big(\psi_{000}\, \psi_{111}\, \psi_{001}\, \psi_{110}+\psi_{000}\, \psi_{111}\, \psi_{010}\, \psi_{101}\\
  &+\psi_{000}\, \psi_{111}\, \psi_{100}\, \psi_{011}+\psi_{001}\, \psi_{110}\, \psi_{010}\, \psi_{101}\\
  \nonumber
  &+\psi_{001}\, \psi_{110}\, \psi_{011}\, \psi_{100}+\psi_{100}\, \psi_{011}\, \psi_{010}\, \psi_{101}\big)\\
  \nonumber
  &+4\,\big(\psi_{000}\, \psi_{011}\, \psi_{101}\, \psi_{110}+\psi_{111}\, \psi_{100}\,  \psi_{010}\, \psi_{001}\big)\,\Big|\,,
\end{aligned}
\end{equation}
where $\psi_{ijk}$ are the coefficients of the state $\ket\psi$ in the computational basis. Since the expression defining the three-tangle is a degree-4 homogeneous polynomial, for any rank-2 state $\rho$ we have in general four complex roots $\{z_i\}_{i=1}^4$ and the pure-state three-tangle in the range of $\rho$ can be therefore expressed as
\begin{equation}
  \tau\Big(\ket{\phi_0} + \omega \ket{\phi_1}\Big) = N_\rho\, \| \bm{\omega} - \bm{z}_1 \|\,  \|\bm{\omega} - \bm{z}_2\|\,\| \bm{\omega} - \bm{z}_3 \|\,\| \bm{\omega} - \bm{z}_4 \|.
\end{equation}
We further remark that using the square root of the three-tangle $\sqrt{\tau}$ as a measure of entanglement, corresponding to $d=4$ and $p=1/2$ in the general expression (\ref{eq:Edp}), is often preferred to using the three-tangle itself due to the simplified SLOCC-invariant properties of polynomial measures with $d p = 2$ for mixed qubit states \cite{viehmann_2012,eltschka_2014,eltschka_2014-1}. However, the geometry of this measure still depends on the four roots of the three-tangle, so we will discuss both cases in the following.

\begin{figure*}[t]
  \centering
  \subfloat[]{\includegraphics[height=5cm]{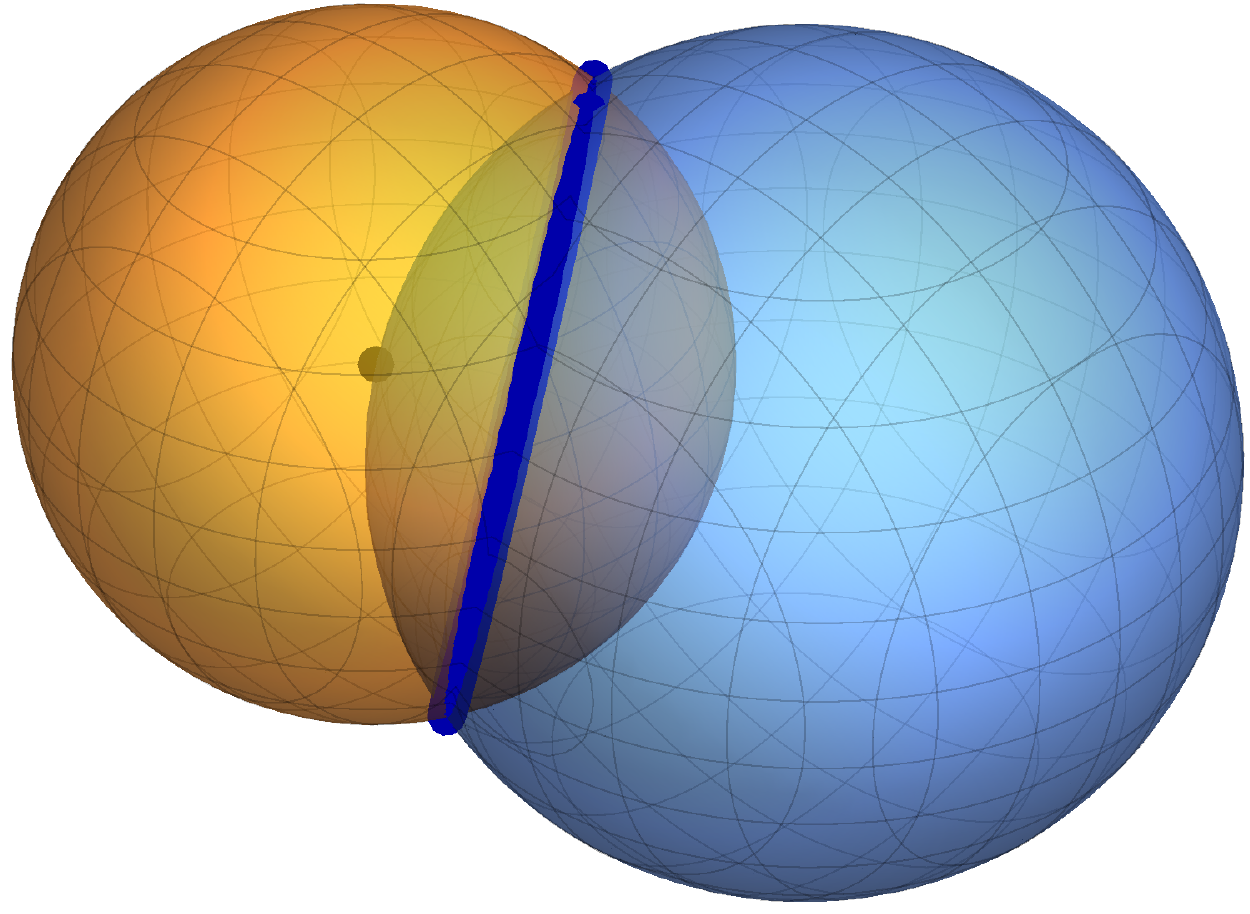}}\hspace*{.3cm}
  \subfloat[]{\includegraphics[height=4.9cm]{conc1.pdf}}\\
  \subfloat[]{\includegraphics[height=5.2cm]{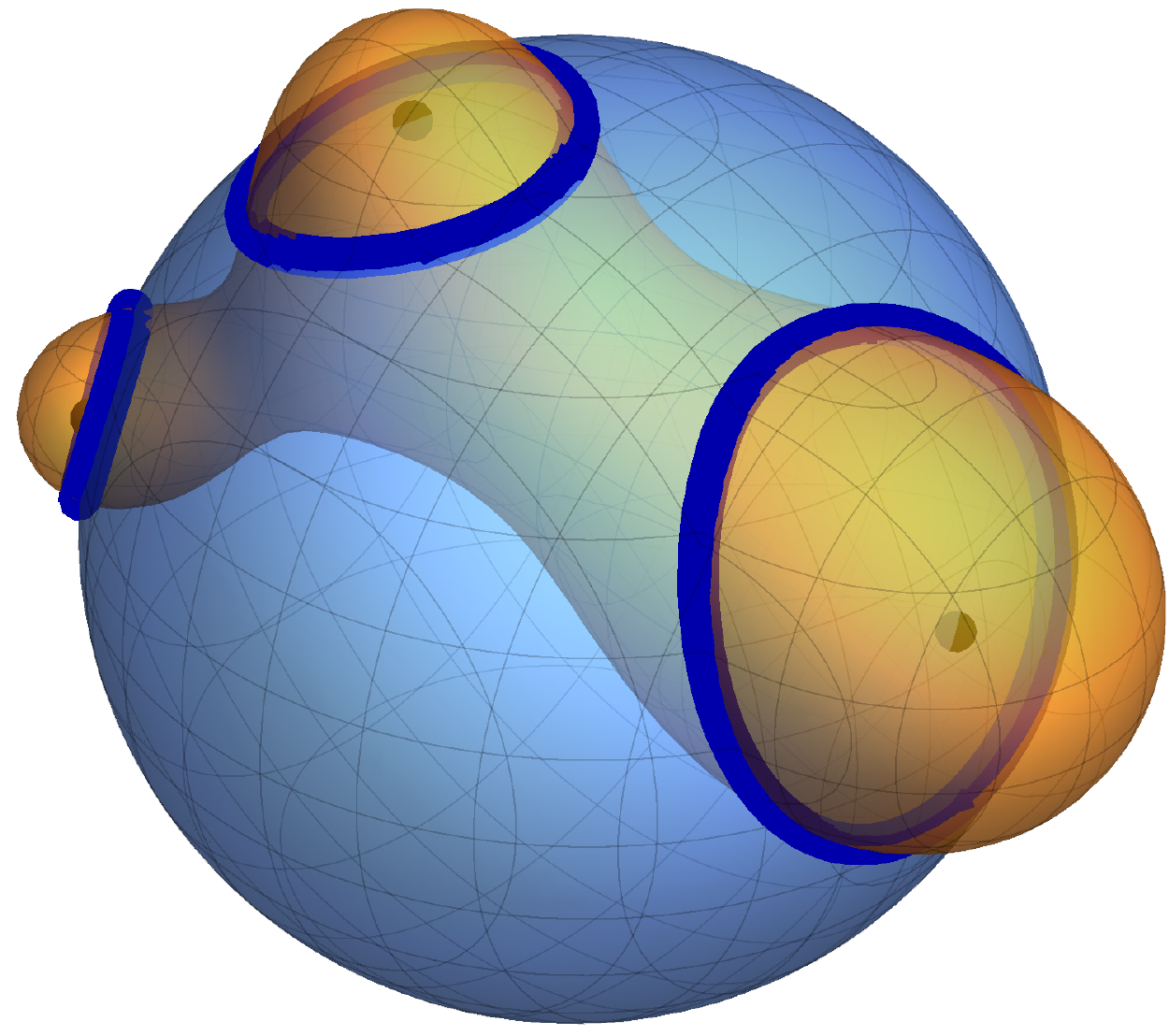}}\hspace*{.3cm}
  \subfloat[]{\includegraphics[height=6cm]{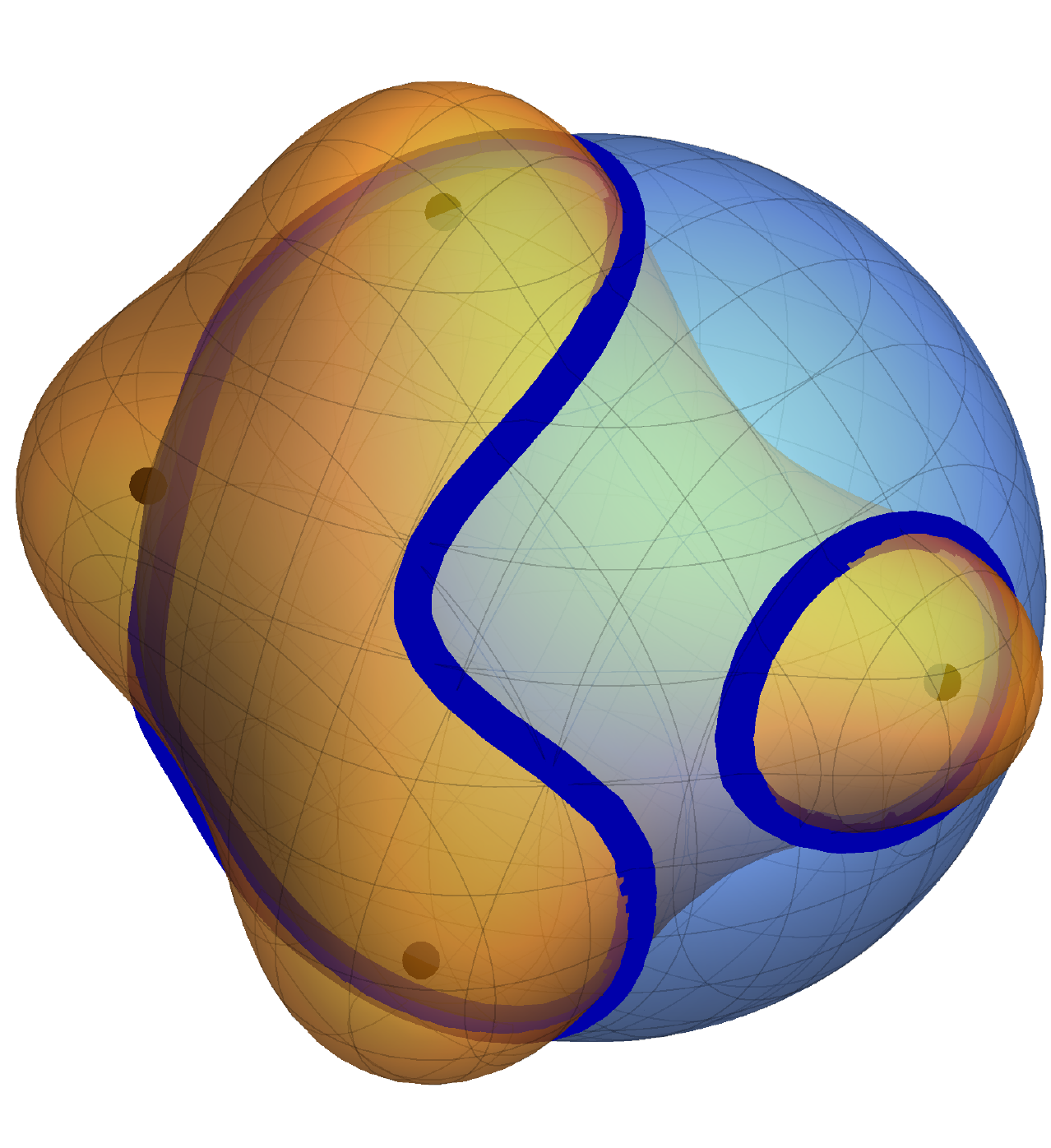}}
  \caption{Visualising curves of constant three-tangle on the Bloch sphere of rank-2 three-qubit states with (a) one, (b) two, (c) three, or (d) four repeating roots. The orange surfaces are defined as the loci of points with a constant product of distances from the foci $\bm z_1, \bm z_2, \bm z_3, \bm z_4$ (the root points). The intersections of the surfaces with the unit sphere represent curves of constant entanglement. The plots show: (a), (c) $\tau = 0.4 N_\rho$, (b), (d) $\tau = N_\rho$, for some choice of roots on the sphere.}
  \label{fig:tangle}
\end{figure*}

We can differentiate several situations which make understanding the curves of constant three-tangle easier for certain states. The simplest case is when we only have one unique root $z_1$ with multiplicity four --- the curves of constant entanglement are then made of points equidistant from the one root point $\bm z_1$, represented by circles on the sphere [Fig.~\ref{fig:tangle}(a)], just as with the two-qubit concurrence. In the case of two unique roots with equal multiplicity (which we will refer to as the {\it two-root} case for simplicity), we regain the peanut-shaped surfaces and the iso-entanglement curves of the concurrence [Fig.~\ref{fig:tangle}(b)]. The other cases do not seem to admit an intuitive explanation for their shape (other than a visual resemblance to ginger roots) and are heavily dependent on the arrangement of the polynomial roots on the Bloch sphere [Fig.~\ref{fig:tangle}(c), (d)].

\subsection{Computable formulas and examples for one and two roots}

We can easily obtain the convex roof of the three-tangle using geometric methods for a state $\rho_1$ with one or a state $\rho_2$ with two repeated roots, since we can see that (in general, for $\rho_2$)
\begin{equation}
  \tau\,\Big(\ket{\phi_0} + \omega \ket{\phi_1}\Big) = N_{\rho_2}\, \big(\| \bm{\omega} - \bm{z}_1 \|\,  \|\bm{\omega} - \bm{z}_2\|\,\big)^2,
\end{equation}
and the geometry of these cases fully corresponds to the case of concurrence discussed in section \ref{sec:concurrence}, so the quantification of the three-tangle effectively reduces to the quantification of the squared concurrence.

\subsubsection{One root}
In the one-root case, we again decompose $\rho_1$ into a convex decomposition that lies on the small circle of the sphere equidistant from the root point, where all of the pure states in the decomposition have the same entanglement. The three-tangle is then given by
\begin{equation}
  \begin{aligned}
    \tau(\rho_1) &= 4\, N_{\rho_1}\, h_c^2\\
    &= 4\, N_{\rho_1} \left| 1 - \braket{z \,|\, \rho_1 |\, z} + \braket{z' \,|\, \rho_1 |\, z'}\right|^2
  \end{aligned}
\end{equation}
following the notation from section \ref{sec:concurrence}. This particular case is simplified further if we take our entanglement measure to be the square root of the three-tangle, as then the measure is of degree 2 and hence is in fact affine throughout the Bloch sphere, taking the same value for any convex decomposition of states in the range of $\rho_1$, just as the concurrence \cite{regula_2016}. Evaluating the convex roof of $\sqrt{\tau}$ is then trivial as no minimization is necessary and all convex decompositions are optimal. Explicit examples can be readily constructed and have been considered in \cite{regula_2016}.

\subsubsection{Two roots}
In the general two-root case, noting that the the calculation of the convex roof extension in section \ref{sec:concurrence} will still apply since the square of a non-negative convex function is still convex \cite{rockafellar_1970}, we simply get
\begin{equation}
  \begin{aligned}
    \tau(\rho_2) &= 4\, N_{\rho_2} \, R^2\, h^2 \\
    &= 4\, N_{\rho_2}\, h^2\, (1-s^2),
  \end{aligned}
\end{equation}
as the convex roof extension of the three-tangle, with the surfaces of constant three-tangle again forming elliptic cylinders inside the Bloch ball, as in Fig.~\ref{fig:concurrence}(b). Following the notation of Fig.~\ref{fig:plot}, a computable formula in terms of density matrices then becomes
\begin{equation}
  \label{eq:tanglerdl}
  \tau\,(\rho_2) = \tau\,\big(\ket{\psi_m}\big)\, \left(\frac{\| \brho - \z_m \|}{\| \bm \psi_m - \z_m \|}\right)^2,
\end{equation}
or simply
\begin{equation}
  \sqrt{\tau}(\rho_2) =  \sqrt{\tau}\big(\ket{\psi_m}\big)\, \frac{\| \brho - \z_m \|}{\| \bm \psi_m - \z_m \|},
\end{equation}
for the square root of the three-tangle. This means, in particular, that any convex decomposition into a (possibly mixed) separable state and a pure state will always be optimal for the the case of the square root of the three-tangle for two-root rank-2 states of three qubits, but not for the three-tangle itself because of the squared factor in Eq.~(\ref{eq:tanglerdl}). Again, example cases of decompositions like this are the best W approximation \cite{acin_2001} or the best zero-$E$ approximation \cite{rodriques_2014}.

All simplified properties detailed in section \ref{sec:concurrence} will also apply here. For instance, for states $\rho_d$ whose two roots are orthogonal to each other, entanglement reduces to coherence in the eigenbasis of the root states, and we get
\begin{align}
  \tau\,(\rho_d) &= 4 N_{\rho_d} \, h^2\nonumber\\
  &= 4\, \tau\,\big(\ket{z'}\big)\,\left| \braket{z_1 \,|\, \rho_d \,|\, z_2}\right|^2,\\
  \sqrt{\tau}\,(\rho_d) &= 2 N_{\rho_d} \, h\nonumber\\
  &= 2 \sqrt{\tau}\big(\ket{z'}\big)\,\left| \braket{z_1 \,|\, \rho_d \,|\, z_2}\right|.\label{eq:sqrttaul1}
\end{align}

As an explicit example, consider the Bloch sphere whose two poles are the generalized W state \cite{coffman_2000}
\begin{equation}
  \ket{W} = a \ket{001} + b \ket{010} + c \ket{100},
\end{equation}
and the generalized flipped W state \cite{jung_2009}
\begin{equation}
  \ket{\widetilde{W}} = d \ket{110} + e \ket{101} + f \ket{011}
\end{equation}
with $a,b,c,d,e,f \in \mathbb{C}$, $|a|^2+|b|^2+|c|^2 = |d|^2+|e|^2+|f|^2 = 1$. The states are explicitly orthogonal, and the three-tangle vanishes on both of them, meaning that convex combinations of these two basis states also have zero three-tangle, and the main (vertical) axis of the sphere will form the zero line. We can then use the geometric approach to quantify the entanglement of any state $\rho_w$ inside this Bloch sphere, as measured by the square root of the three-tangle, giving
\begin{equation}
  \begin{aligned}
    \sqrt\tau (\rho_w) = &2 \sqrt{\left| a^2 d^2+b^2 e^2+c^2 f^2-2 (a b d e + a c d f+ b c e f)\right|}\\
    &\times \left|\braket{W \,|\, \rho_w \,|\, \widetilde{W}}\right|,
  \end{aligned}
\end{equation}
where we have explicitly calculated the normalization constant appearing in Eq.~(\ref{eq:sqrttaul1}).

\subsection{Progress for three and four roots}
To investigate the case of three or four different roots, we can follow in principle the same idea. By constructing the convex hull of the curves of constant three-tangle for a particular value of $\tau=\tau_0$, we obtain the surface consisting of mixed states $\rho$ which can be decomposed into pure states all having $\tau=\tau_0$, and the smallest such decomposition (the surface closest to the zero polytope) is a candidate for the optimal decomposition of $\rho$. However, determining when such a decomposition is in fact optimal still remains a non-trivial task, and there does not appear to be a straightforward solution to the problem in the general case. One reason why it has not been possible so far to quantify the convex roof in these more complex situations, geometrically or otherwise, is the loss of the Bloch sphere symmetries that we could exploit in the one- and two-root cases. Such symmetries seem to be a crucial ingredient in analysing the properties of the convex roof extension. One can then look at special cases when the situation is indeed symmetric.

\subsubsection{Mixtures of GHZ and W states}

The Greenberger-Horne-Zeilinger (GHZ) and W states represent two fundamental and inequivalent kinds of three-qubit entanglement \cite{dur_2000}. The solution of the convex roof problem for their mixtures of the form
\begin{equation}
  \rho(p) = p \ket{\text{GHZ}}\bra{\text{GHZ}} + (1-p) \ket{\text{W}}\bra{\text{W}},
\end{equation}
with $0 \leq p \leq 1$,
was first obtained by Lohmayer et al. \cite{lohmayer_2006} and, along with its generalizations \cite{eltschka_2008,jung_2009,viehmann_2012}, still constitutes one of the few cases where the exact quantification of mixed-state three-tangle is available. We will show that a procedure analogous to the one in \cite{lohmayer_2006} can be carried out in purely geometric terms. To this aim, we can notice strong symmetries in the Bloch sphere: the three-tangle has four roots, but one of them is a pole of the sphere (the W state) and the other three are the vertices of an equilateral triangle parallel to the equatorial plane (see e.g.~the figures and the explicit derivations in \cite{lohmayer_2006,osterloh_2008,eltschka_2008}).

\begin{figure*}[t]
  \centering
  \subfloat[]{\includegraphics[width=6cm]{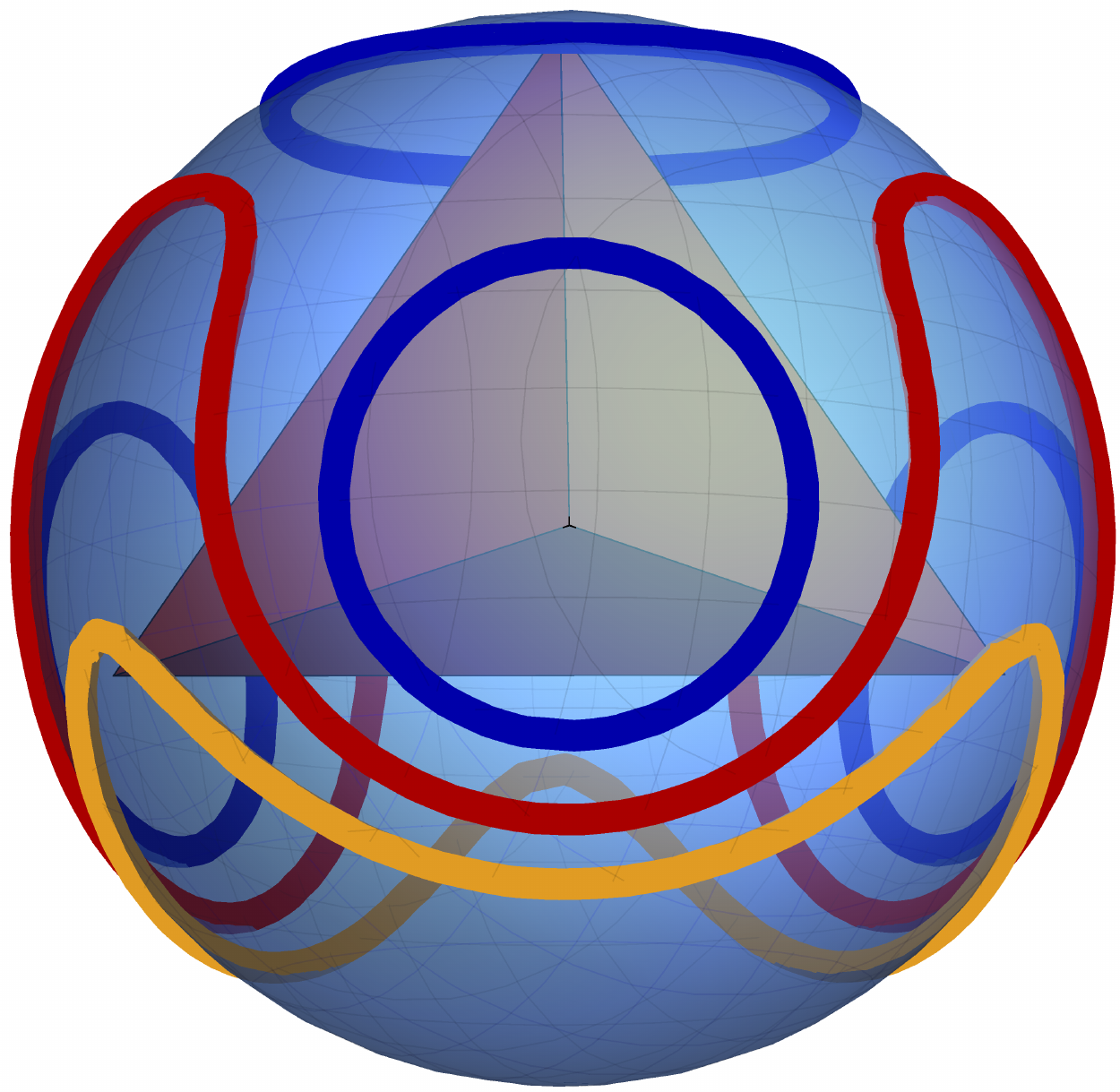}}\hspace*{.3cm}
  \subfloat[]{\includegraphics[width=8.5cm]{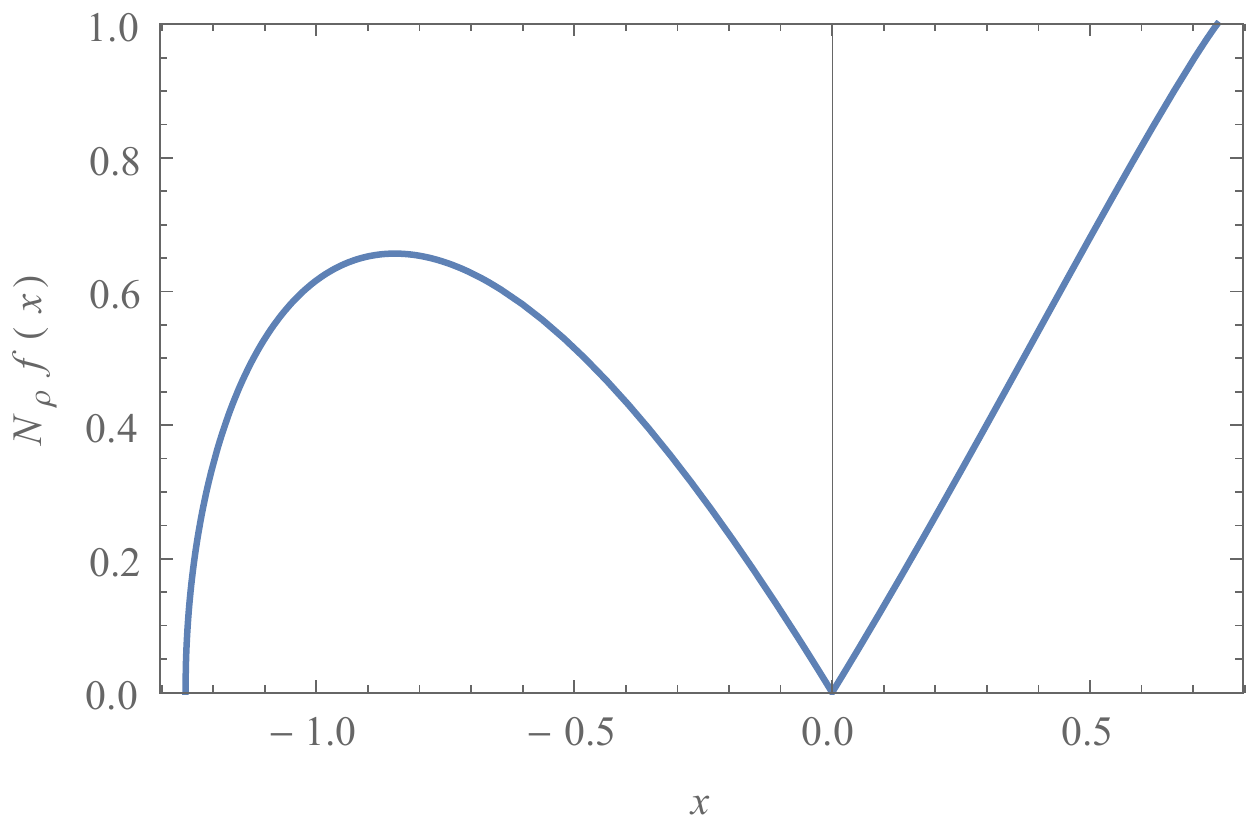}}
  \caption{(a) The radial spread of the curves of constant three-tangle from the zero simplex of GHZ and W mixtures (shaded pyramid). The curves are plotted for $\tau = \sqrt{3} N_\rho$ (blue curve), $\tau = \sqrt{6} N_\rho$ (red curve) and $\tau = 3 N_\rho$ (yellow curve). We note that the Bloch sphere in our notation is upside-down with respect to the one in \cite{lohmayer_2006}.\newline\hspace{\textwidth}
    (b) The function $f(x)$ for the mixtures of GHZ and W states, with the normalization constant $N_\rho$ restored. We note that for $x<0$ this corresponds to the inside of the zero simplex, so the three-tangle in that region has to vanish. $f(x)$ stays convex until $x \approx 0.14$, after which a different decomposition of the state $\rho$ is necessary.}
  \label{fig:ghz}
\end{figure*}

Let us analyse this case geometrically. We have a zero simplex $\z_1 \z_2 \z_3 \z_4$ whose base is an equilateral triangle $\z_1 \z_2 \z_3$ and the point $\z_4$ lies at the north pole of the Bloch sphere (corresponding in our notation to the W state). We then want to quantify the entanglement of states with Bloch vectors $\brho_c$ lying on the main axis of the sphere, by choosing a suitable decomposition into states $\{\bpsi_i\}$ on the surface of the sphere. Due to the symmetry of the problem, all pure states lying on planes perpendicular to the main axis will be equidistant from $\z_4$.  We can then choose three points $\bpsi_1,\bpsi_2,\bpsi_3$ in the plane containing $\brho_c$ and perpendicular to the main axis for which the product $P(\bpsi_i | \z_1 \z_2 \z_3)$ will be the same: if the three points form the vertices of another equilateral triangle in the plane, then the three tetrahedra $\bpsi_1 \z_1 \z_2 \z_3$, $\bpsi_1 \z_1 \z_2 \z_3$, and $\bpsi_1 \z_1 \z_2 \z_3$ will be identical and so any such triangle in the plane defines a constant decomposition of the state $\rho_c$. Out of these triangles, we choose the one which minimizes the product of distances: seeing as the curves of constant three-tangle spread out radially from each of the root points [Fig.~\ref{fig:ghz}(a)], such a triangle can be obtained by choosing vertices $\bpsi_1,\bpsi_2,\bpsi_3$ lying along the meridians through $\z_1, \z_2, \z_3$.

To obtain a function which to every point $\bm\rho_c$ inside the ball assigns the value of the product $P$ calculated at one of the vertices $\bpsi_1,\bpsi_2,\bpsi_3$ from its decomposition, we employ some elementary geometry, in particular the Pythagorean theorem applied to different triangles in this construction. This gives
\begin{equation}
  \begin{aligned}
    f(\rho_c) &= P(\bpsi_i \,|\, \z_1 \z_2 \z_3 \z_4)\\
    &= \sqrt{h^2 + (R_{\psi_1 \psi_2 \psi_3}-R_{z_1 z_2 z_3})^2}\,\sqrt{H^2 + R_{\psi_1 \psi_2 \psi_3}^2}\\
    &\;\times\,\left(h^2+R_{\psi_1 \psi_2 \psi_3}^2+R_{z_1 z_2 z_3}^2+R_{z_1 z_2 z_3}R_{\psi_1 \psi_2 \psi_3}\right)
  \end{aligned}
\end{equation}
where $h$ is the distance of $\brho_c$ to the zero simplex (i.e. to the plane containing $\triangle \z_1 \z_2 \z_3$), $H$ is the distance of $\brho_c$ to $\z_4$ (pole), and $R$ denotes the circumradius of the corresponding triangle (the radius of the small circle of the sphere which contains it). This clearly holds for any point $\bm \rho \in \triangle \bpsi_1 \bpsi_2 \bpsi_3$, so any state inside of it has the same constant decomposition. Since we are interested in the points $\brho_c$ on the main axis of the sphere, let us write $f$ in one coordinate $x$ centered at the point of intersection of the main axis with $\triangle \z_1 \z_2 \z_3$ and increasing away from $\z_4$. We then get
\begin{equation}
  \begin{aligned}
    f(x) = &\,2 \sqrt{x-x_O+1}\\
    & \times \sqrt{1+x_O (x-x_O)-\sqrt{\left(x_O^2-1\right) \left[(x-x_O)^2-1\right]}}\\
    & \times \left(\sqrt{\left(x_O^2-1\right) \left[(x-x_O)^2-1\right]}+2 x_O (x-x_O)+2\right)	
  \end{aligned}
\end{equation}
where $x_O$ is the coordinate of the origin of the sphere in these coordinates. Since the entanglement has to vanish for all states inside of the zero simplex ($x\leq0$), we are interested in values of this function for positive $x$. However, by explicitly evaluating the second derivative of $f(x)$, we see that this function is convex only in an interval $0 \leq x \leq x'$ for a value of $x'$ depending on $x_O$, which means that this decomposition is in fact not optimal for all states on the axis.

Going back to the specific case of GHZ and W mixtures, this setting has $x_O = 6/\left(3+4\sqrt[3]{2}\right) - 1$ and the function $N_\rho f(x)$, as expected, corresponds to $g_{I}(p)$ in the notation of Ref.~\cite{lohmayer_2006}, see Fig.~\ref{fig:ghz}(b). This function is convex up to a point $x' \approx 0.14$, meaning that the flat decomposition is indeed optimal in this region, but a different decomposition is necessary in the region $x \geq x'$. In fact, it has been shown that the convex roof of the three-tangle is given by the convex hull of the function $N_\rho f(x)$ \cite{osterloh_2008,eltschka_2008}, and an analogous derivation of this fact can be performed geometrically following the same steps. We note, however, that there is no \emph{a priori} reason why this should hold in more general states which do not enjoy similar geometric symmetries.

This shows that the geometric approach can indeed be useful in understanding and quantifying the entanglement in more complicated situations, although the constant decomposition which was proven optimal in the one- and two-root cases is no longer always optimal for all four-root states and additional measures have to be taken to account for that.

\subsection{Classification of rank-2 three-qubit states}

We have shown that our geometric method allows us to quantify the convex roof of the three-tangle (or its square root) exactly in the specific cases of one- and two-root states. One might then wonder what kinds of states obey these conditions, and how common they are among all three-qubit states. To investigate this, we note that any rank-2 three-qubit state can be obtained from a pure four-qubit state by tracing out one qubit, allowing us to look at the set of all four-qubit pure states first.

A common way to classify states with regards to their entanglement properties is to consider the equivalence classes generated by SLOCC transformations \cite{dur_2000}, that is, to divide the set of all states into classes of states which cannot be transformed into each other by SLOCC. In fact, while three-qubit states have only two such SLOCC-inequivalent classes (represented respectively by GHZ and W states), four qubits have infinitely many such classes \cite{dur_2000}. However, there does exist a way to group the SLOCC classes of four-qubit states into nine families, each representing a fundamentally different type of entanglement \cite{verstraete_2002, chterental_2007, spee_2015}. Each of these families can be represented by a generating family of states $\ket{G_{abcd}^{\,\mu}}$ where $\mu \in \{1, \ldots, 9\}$ and $a, b, c, d$ are complex parameters with non-negative real part. The states $\ket{G_{abcd}^{\,\mu}}$ are defined (up to normalization) as follows:
\begin{eqnarray}
  \ket{G^1_{abcd}}&=&\frac{a+d}{2}\big(\ket{0000}+\ket{1111}\big)+\frac{a-d}{2}\big(\ket{0011}+\ket{1100}\big)\nonumber\\
  && \hspace*{-5pt}+\frac{b+c}{2}\big(\ket{0101}+\ket{1010}\big) + \frac{b-c}{2}\big(\ket{0110}+\ket{1001}\big),\nonumber\\
  \ket{G^2_{abc}} &=& \frac{a+b}{2}\big(\ket{0000}+\ket{1111}\big)+\frac{a-b}{2}\big(\ket{0011}+\ket{1100}\big)\nonumber \\
  &&+ c\,\big(\ket{0101}+\ket{1010}\big) + \ket{0110},\nonumber\\
  \ket{G^3_{ab}}&=&a\,\big(\ket{0000}+\ket{1111}\big) + b\,\big(\ket{0101}+\ket{1010}\big)\nonumber\\
  &&+ \ket{0110} + \ket{0011},\nonumber\\
  \ket{G^4_{ab}}&=&a\,\big(\ket{0000}+\ket{1111}\big) + \frac{a+b}{2}\big(\ket{0101}+\ket{1010}\big)\nonumber\\
  &&+ \frac{a-b}{2}\big(\ket{0110}+\ket{1001}\big)\\
  &&+ \frac{i}{\sqrt{2}}\big(-\ket{0001} - \ket{0010} + \ket{0111} + \ket{1011}\big),\nonumber\\
  \ket{G^5_a}&=&  a\,\big(\ket{0000} + \ket{0101} + \ket{1010} + \ket{1111}\big)\nonumber\\
  && +i\ket{0001} + \ket{0110} - i\ket{1011},\nonumber\\
  \ket{G^6_a}&=& a\,\big(\ket{0000} + \ket{1111}\big) + \ket{0011} + \ket{0101} + \ket{0110},\nonumber\\
  \ket{G^7}&=& \ket{0000} + \ket{0101} + \ket{1000} + \ket{1110},\nonumber\\
  \ket{G^8}&=& \ket{0000} + \ket{1011} + \ket{1101} + \ket{1110},\nonumber\\
  \ket{G^9}&=& \ket{0000} + \ket{0111},\nonumber
\end{eqnarray}
where we note that the formula for the generating state $\ket{G_{ab}^4}$ was reported incorrectly in the original classification of Ref.~\cite{verstraete_2002} and corrected later in Ref.~\cite{chterental_2007} (see also \cite{spee_2015}).

All states belonging to the $\mu$th class can then be constructed as
\begin{equation}
  \ket{\Psi^\mu} = \frac{L \ket{G_{abcd}^{\,\mu}}}%
      {\| L \ket{G_{abcd}^{\,\mu}} \|},
\end{equation}
where
\begin{equation}
  L = A_1 \otimes A_2 \otimes A_3 \otimes A_4,
\end{equation}
with $A_i \in \text{SL}(2,\mathbb{C})$ denoting SLOCC transformations on each qubit. Since the union of all nine classes spans the Hilbert space of all four-qubit pure states \cite{verstraete_2002}, we can then obtain any rank-2 three-qubit state as
\begin{equation}
  \rho = \Tr_k \left[ \frac{ L \ket{G_{abcd}^{\,\mu}} \bra{G_{abcd}^{\,\mu}} L^{\dagger} } { \Tr \left( L \ket{G_{abcd}^{\,\mu}} \bra{G_{abcd}^{\,\mu}} L^{\dagger} \right) } \right]
\end{equation}
for some choice of SLOCC transformations $L$ in the class $\mu$ and some values of $a,b,c,d$. Here, $k \in \{1,2,3,4\}$ denotes which qubit is traced out from the four-qubit state.

To proceed with the classification of three-qubit states, we note that since SLOCC operations on four-qubit states leave the number of pure states with zero three-tangle in their reduced subsystems (i.e., the number of roots in the corresponding zero polytopes) invariant \cite{regula_2016}, it is sufficient to investigate the generating families $\ket{G_{abcd}^{\,\mu}}$ to obtain the number of three-tangle roots for the three-qubit marginals of any state in the corresponding classes. We also note that particular choices of the parameters $a,b,c,d$ can lead to degeneracies in the SLOCC classes \cite{djokovic_2009,spee_2015}, and so we classify these degenerate subclasses separately if the degeneracy leads to a different number of roots. We then verify the number of polynomial roots of the three-tangle in the linear combinations of eigenvectors for each reduced marginal.

We report our results in Table \ref{table:slocc}. Remarkably, the marginals of several classes of states have only one or two roots (with equal multiplicites), which means that the general geometric methods developed in this paper allow for the exact quantification of the three-tangle of mixed states obtained from a significant range of different types of entangled four-qubit states. These properties can be seen to be particularly common in the degenerate subclasses of the SLOCC classification.

An important incentive for quantifying the entanglement exactly in the reduced subsystems of the generating states lies in the fact that if one employs an entanglement measure of degree $dp=2$, the entanglement of a given state $\rho$ and the entanglement of any state $L\rho L^\dagger$ obtained from it by a SLOCC transformation are related by a simple linear scaling, analogous to the case of polynomial invariants for pure states \cite{viehmann_2012}. For example, for the square root of the three-tangle we have
\begin{equation}
  \sqrt\tau \left( \frac{L\rho L^\dagger}{\Tr \left(L\rho L^\dagger\right)} \right) = \frac{\sqrt\tau\left(L\rho L^\dagger\right)}{\Tr \left(L\rho L^\dagger\right)}.
\end{equation}
This means that, in the case of $\sqrt{\tau}$, our results yield computable methods for the exact determination of the entanglement in {\it all} three-qubit rank-2 states generated via arbitrary SLOCC transformations applied to any representative state in the SLOCC classification of Table~\ref{table:slocc} with one or two roots (with equal multiplicities). We also remark that given any four-qubit pure state it is possible to identify the generating family of the state's SLOCC class by following the methods and algorithms given in Refs. \cite{holweck_2014,holweck_2016}.

\begin{table*}[t]
  \centering
  \caption{The number of roots of the three-tangle for the linear combination of eigenvectors of reduced subsystems of the nine families of four-qubit pure states and their degenerate subclasses. $2^{*}$ denotes a state which has two roots, but they have unequal multiplicities. ``Pure'' denotes reduced subsystems which are actually rank 1. Degenerate subclasses not included in the table have the same number of roots as the generic subclass.}
  \label{table:slocc}
  \begin{tabular}{@{}cl cccc}
    \toprule[0.2pt]
    \multirow{2}{*}{$\;$Class$\;$} & \multirow{2}{*}{$\;\;$Subclass$\;\;$} & \multicolumn{4}{c}{Number of roots in} \\
    &  & \multicolumn{1}{c}{$\text{Tr}_1 \ket{G^{\mu}}\bra{G^{\mu}}$} & \multicolumn{1}{c}{$\text{Tr}_2 \ket{G^{\mu}}\bra{G^{\mu}}$} & \multicolumn{1}{c}{$\text{Tr}_3 \ket{G^{\mu}}\bra{G^{\mu}}$} & \multicolumn{1}{c}{$\text{Tr}_4 \ket{G^{\mu}}\bra{G^{\mu}}$} \\
    \midrule[0.1pt]
    \multicolumn{1}{c}{$\ket{G_{abcd}^1}$} & \textbf{Generic} & \multicolumn{4}{c}{\textbf{4}} \\
    \cline{2-6}
    & \begin{tabular}[c]{@{}l@{}}$a=\pm\, b$\\ $a=\pm\, c$\\ $a=\pm\, d$\\ $b=\pm\, c$\\$b=\pm\, d$\\$c=\pm\, d$\end{tabular} & \multicolumn{4}{c}{2} \\
    \cline{2-6}
    & \begin{tabular}[c]{@{}l@{}}$a=\pm\, b=\pm\, c$\\ $a=\pm\, b=\pm\, d$\\ $a=\pm\, c=\pm\, d$\\ $b=\pm\, c=\pm\, d$\end{tabular} & \multicolumn{4}{c}{4} \\
    \midrule[0.1pt]
    $\ket{G_{abc}^2}$ & \textbf{Generic} & \multicolumn{4}{c}{\textbf{3}} \\
    \cline{2-6}
    & \begin{tabular}[c]{@{}l@{}}$a=\pm\, b$\\ $c=0$\end{tabular} & \multicolumn{4}{c}{2} \\
    \cline{2-6}
    & \begin{tabular}[c]{@{}l  @{}}$a=\pm\, c$\\ $b=\pm\, c$\end{tabular} & \multicolumn{4}{c}{1} \\
    \cline{2-6}
    & \begin{tabular}[c]{@{}l  @{}}$a=\pm\, b=\pm\, c$\\$a=c=0$\end{tabular} & \multicolumn{4}{c}{4} \\
    \cline{2-6}
    & \begin{tabular}[c]{@{}l  @{}}$a=b=c=0$\end{tabular} & \multicolumn{4}{c}{pure} \\
    \midrule[0.1pt]
    $\ket{G_{ab}^3}$ & \textbf{Generic} & \textbf{3} & \textbf{2} & \textbf{3} & \textbf{2} \\
    \cline{2-6}
    & $a=\pm\, b$ & 1 & 4 & 1 & 4 \\
    \cline{2-6}
    & \begin{tabular}[c]{@{}l@{}}$a=0$\\ $b=0$\end{tabular} & \multicolumn{4}{c}{2} \\
    \cline{2-6}
    & $a=b=0$ & pure & 4 & pure & 4 \\
    \midrule[0.1pt]
    $\ket{G_{ab}^4}$ & \textbf{Generic} & \multicolumn{4}{c}{\textbf{2}$^{*}$} \\
    \cline{2-6}
    & \begin{tabular}[c]{@{}l@{}}$a=\pm\, b$\\ $a=0$\\ $b=0$\end{tabular} & \multicolumn{4}{c}{1} \\
    \cline{2-6}
    & \begin{tabular}[c]{@{}l@{}}$a=b=0$\end{tabular} & \multicolumn{4}{c}{4} \\
    \midrule[0.1pt]
    $\ket{G_{a}^5}$ & \textbf{Generic} & \textbf{2}$^{*}$ & \textbf{1} & \textbf{2}$^{*}$ & \textbf{1} \\
    \cline{2-6}
    & $a=0$ & 1 & 4 & 1 & 4 \\
    \midrule[0.1pt]
    $\ket{G^6_{a}}$ & \textbf{Generic} & \textbf{3} & \textbf{2} & \textbf{2} & \textbf{2} \\
    \cline{2-6}
    & $a=0$ & pure & 4 & 4 & 4\\
    \midrule[0.1pt]
    $\ket{G^7}$ & \textbf{Generic} & \textbf{4} & \textbf{1} & \textbf{1} & \textbf{1} \\
    \midrule[0.1pt]
    $\ket{G^8}$ & \textbf{Generic} & \textbf{2}$^{*}$& \textbf{1} & \textbf{1} & \textbf{1} \\
    \midrule[0.1pt]
    $\ket{G^9}$ & \textbf{Generic} & pure & \textbf{4} & \textbf{4} & \textbf{4}\\
    \bottomrule[0.2pt]
  \end{tabular}
\end{table*}



\section{General polynomial measures}\label{sec:general}

The cases considered in this paper can be straightforwardly generalized to any entanglement measure based on a polynomial invariant of an even degree $d$: if a given state has only one or two roots with equal multiplicities, one can employ the geometric formulas for the concurrence to the $\frac{d}{2}$th power. For multiqubit states, noting that there are no nontrivial invariants of odd degree \cite{brylinski_2002}, this applies to all polynomial invariants of interest. Explicitly, we have:
\begin{align}
  E_d\,(\rho_1) &= N_{\rho_1} \left( 2\,h_c \right)^{\,d/2}\nonumber\\
  &= E_d\big(\ket{z'}\big) \left( \frac{1}{2} \left|1 - \braket{z \,|\, \rho_1 |\, z} + \braket{z' \,|\, \rho_1 |\, z'}\right|\right)^{\,d/2},\\
  E_d\,(\rho_2) &= N_{\rho_2} \left( 2\,R\,h \right)^{\,d/2}\nonumber\\
  &= E_d \big(\ket{\,\psi_m}\big)\, \left(\frac{\| \brho - \z_m \|}{\| \bm \psi_m - \z_m \|}\right)^{\,d/2},
\end{align}
in the notation of sections \ref{sec:concurrence} and \ref{sec:tangle}, for any polynomial entanglement measure $E_d$ and all rank-2 states with one ($\rho_1$) or two ($\rho_2$) unentangled states in their range. All other simplified properties and formulas discussed before still apply as well.

More generally, the quantification of any polynomial measure with repeated roots will be reduced to the quantification of a lower-degree measure --- for instance, if a measure based on a degree-8 polynomial has 4 repeated roots whose geometry corresponds to a known solution for the three-tangle, we can quantify its convex roof exactly using the solutions obtained in the three-tangle case, etc. The geometric approach itself is of course valid for any number of roots and any degree of the polynomial measure, and may come in particularly handy when one can exploit geometric symmetries of the problem, as we have demonstrated by analysing the entanglement of mixtures of GHZ and W states of three qubits.

We further note that, since entanglement measures of homogeneous degree 2 are particularly useful thanks to their simplified SLOCC rescaling properties \cite{viehmann_2012,eltschka_2014-1}, one might consider what happens when we take $E_d^{2/d}$ as our measure, i.e., when we set $p=2/d$ in Eq.~(\ref{eq:Edp}). The formula for the entanglement of a two-root state then becomes
\begin{equation}
  E_d^{2/d} (\rho_2) = [E_d\big(\ket{\,\psi_m}\big)]^{2/d}\, \frac{\| \brho - \z_m \|}{\| \bm \psi_m - \z_m \|},
\end{equation}
and we regain many of the even simpler and linear properties of the concurrence, in particular the property that any decomposition into one (possibly mixed) separable state and one pure state is always optimal (see section \ref{sec:concurrence}). As a result, bounds obtained from the best zero-$E$ approximation \cite{rodriques_2014} are tight for all two-root multiqubit states with no optimization required. In the case of one-root states, measures of degree 2 enjoy the even stronger property that every convex decomposition of a mixed state is in fact optimal, and the problem of the convex roof becomes trivial \cite{regula_2016}. This provides further evidence for the privileged position of degree 2 among all possible homogeneous degrees for polynomial entanglement measures \cite{viehmann_2012}.

\section{Conclusions}\label{sec:concl}

We have shown that the quantification of polynomial measures of entanglement for rank-2 states of multipartite systems can be understood in geometric terms, and in many relevant cases these methods provide novel insights into the properties of such states as well as computable formulas for their entanglement. In particular, for pure states in the range of a rank-2 mixed state (geometrically spanning a Bloch sphere), the quantification of any entanglement monotone based on a polynomial invariant of homogeneous degree $d$ corresponds simply to measuring the product of Euclidean distances between the state's Bloch vector and the $d$ unentangled states (polynomial roots) on the surface of the Bloch sphere. This can then be used to obtain a more intuitive visual representation of the entanglement structure, providing novel efficient ways to understand and quantify the convex roof extension of all such polynomial entanglement measures to the set of mixed states.

We have explicitly demonstrated that the convex roof of the concurrence of two qubits \cite{hill_1997,wootters_1998} can be reobtained by relying only on the geometric approach, and we have shown that the quantification of the convex roof of any polynomial measure of entanglement, such as the three-tangle (or its square root) for three qubits, is effectively reduced to the case of the concurrence if the state whose entanglement is being computed has only one or two unentangled roots in its range. Additionally, we have shown that such one- and two-root states in fact appear as the marginals of several classes of four-qubit pure states.

The exact quantification of the convex roof in the one- and two-root marginals of four-qubit states can help us gain a better understanding of their entanglement distribution properties, such as the monogamy relations \cite{coffman_2000}, regarded as a fundamental property of quantum entanglement \cite{terhal_2004,streltsov_2012,eltschka_2015,lancien_2016}. For instance, rather surprisingly, the one-root degenerate subclass of class $\ket{G_{abc}^{\,2}}$ (see Table~\ref{table:slocc}) yields one rare instance of four-qubit states which violate a seemingly natural generalization of the Coffman-Kundu-Wootters monogamy inequality \cite{coffman_2000, osborne_2006} where multipartite entanglement is considered in addition to the bipartite terms \cite{regula_2014,regula_2016-1}.

The possibility of further generalising the methods presented here is open, but not straightforward. While the polynomial invariants for larger systems can be obtained \cite{osterloh_2005,gour_2013} and the geometric methods themselves could be extended to a generalized Bloch vector formalism \cite{kimura_2003},  states of a higher rank do not, in general, admit a finite number of polynomial solutions on the set of pure states. This makes a direct application of our methods unfeasible, although constructing an analogous approach for higher-rank states could certainly be a possible extension of this work. Another direction of future research would be to find classes of $n$-qubit or $n$-qudit states whose properties allow for a simplified quantification of their entanglement based on special geometric features, analogous to one- and two-root states.

We hope that our methods can find their use in understanding and characterising the intricate properties of multipartite entanglement, and that the geometric insight they provide can be successfully applied to quantify polynomial measures of entanglement exactly in a wider variety of quantum states. In particular, obtaining a closed formula for the (square root of) three-tangle for all rank-2 states of three qubits, going beyond the one-root case solved in \cite{regula_2016} and the two-root case solved here, would be a remarkable achievement.

\section*{Acknowledgements}
We thank the European Research Council (ERC) Starting Grant GQCOP (Grant No.~637352), for financial support. We are grateful to Sooji Han and Fedor Petrov for help with the mathematical content of the paper, and to Jens Siewert and Ajit Iqbal Singh for their careful comments on drafts of this manuscript. We further acknowledge insightful discussions with Thomas R. Bromley, Marco Cianciaruso, Antony Milne, Zbigniew Pucha\l a, Jens Siewert, Alexander Streltsov, and Karol \.{Z}yczkowski.

\bibliographystyle{apsrev4-1}
\footnotesize
\bibliography{main}

\begin{thebibliography}{61}%
\makeatletter
\providecommand \@ifxundefined [1]{%
 \@ifx{#1\undefined}
}%
\providecommand \@ifnum [1]{%
 \ifnum #1\expandafter \@firstoftwo
 \else \expandafter \@secondoftwo
 \fi
}%
\providecommand \@ifx [1]{%
 \ifx #1\expandafter \@firstoftwo
 \else \expandafter \@secondoftwo
 \fi
}%
\providecommand \natexlab [1]{#1}%
\providecommand \enquote  [1]{``#1''}%
\providecommand \bibnamefont  [1]{#1}%
\providecommand \bibfnamefont [1]{#1}%
\providecommand \citenamefont [1]{#1}%
\providecommand \href@noop [0]{\@secondoftwo}%
\providecommand \href [0]{\begingroup \@sanitize@url \@href}%
\providecommand \@href[1]{\@@startlink{#1}\@@href}%
\providecommand \@@href[1]{\endgroup#1\@@endlink}%
\providecommand \@sanitize@url [0]{\catcode `\\12\catcode `\$12\catcode
  `\&12\catcode `\#12\catcode `\^12\catcode `\_12\catcode `\%12\relax}%
\providecommand \@@startlink[1]{}%
\providecommand \@@endlink[0]{}%
\providecommand \url  [0]{\begingroup\@sanitize@url \@url }%
\providecommand \@url [1]{\endgroup\@href {#1}{\urlprefix }}%
\providecommand \urlprefix  [0]{URL }%
\providecommand \Eprint [0]{\href }%
\providecommand \doibase [0]{http://dx.doi.org/}%
\providecommand \selectlanguage [0]{\@gobble}%
\providecommand \bibinfo  [0]{\@secondoftwo}%
\providecommand \bibfield  [0]{\@secondoftwo}%
\providecommand \translation [1]{[#1]}%
\providecommand \BibitemOpen [0]{}%
\providecommand \bibitemStop [0]{}%
\providecommand \bibitemNoStop [0]{.\EOS\space}%
\providecommand \EOS [0]{\spacefactor3000\relax}%
\providecommand \BibitemShut  [1]{\csname bibitem#1\endcsname}%
\let\auto@bib@innerbib\@empty
\bibitem [{\citenamefont {Plenio}\ and\ \citenamefont
  {Virmani}(2007)}]{plenio_2007}%
  \BibitemOpen
  \bibfield  {author} {\bibinfo {author} {\bibfnamefont {M.~B.}\ \bibnamefont
  {Plenio}}\ and\ \bibinfo {author} {\bibfnamefont {S.}~\bibnamefont
  {Virmani}},\ }\href@noop {} {\bibfield  {journal} {\bibinfo  {journal}
  {Quant. Inf. Comp.}\ }\textbf {\bibinfo {volume} {7}},\ \bibinfo {pages} {1}
  (\bibinfo {year} {2007})}\BibitemShut {NoStop}%
\bibitem [{\citenamefont {Horodecki}\ \emph {et~al.}(2009)\citenamefont
  {Horodecki}, \citenamefont {Horodecki}, \citenamefont {Horodecki},\ and\
  \citenamefont {Horodecki}}]{horodecki_2009}%
  \BibitemOpen
  \bibfield  {author} {\bibinfo {author} {\bibfnamefont {R.}~\bibnamefont
  {Horodecki}}, \bibinfo {author} {\bibfnamefont {P.}~\bibnamefont
  {Horodecki}}, \bibinfo {author} {\bibfnamefont {M.}~\bibnamefont
  {Horodecki}}, \ and\ \bibinfo {author} {\bibfnamefont {K.}~\bibnamefont
  {Horodecki}},\ }\href {\doibase 10.1103/RevModPhys.81.865} {\bibfield
  {journal} {\bibinfo  {journal} {Rev Mod Phys}\ }\textbf {\bibinfo {volume}
  {81}},\ \bibinfo {pages} {865} (\bibinfo {year} {2009})}\BibitemShut
  {NoStop}%
\bibitem [{\citenamefont {Eltschka}\ and\ \citenamefont
  {Siewert}(2014{\natexlab{a}})}]{eltschka_2014-1}%
  \BibitemOpen
  \bibfield  {author} {\bibinfo {author} {\bibfnamefont {C.}~\bibnamefont
  {Eltschka}}\ and\ \bibinfo {author} {\bibfnamefont {J.}~\bibnamefont
  {Siewert}},\ }\href {\doibase 10.1088/1751-8113/47/42/424005} {\bibfield
  {journal} {\bibinfo  {journal} {J. Phys. A: Math. Theor.}\ }\textbf {\bibinfo
  {volume} {47}},\ \bibinfo {pages} {424005} (\bibinfo {year}
  {2014}{\natexlab{a}})}\BibitemShut {NoStop}%
\bibitem [{\citenamefont {Vedral}\ \emph {et~al.}(1997)\citenamefont {Vedral},
  \citenamefont {Plenio}, \citenamefont {Rippin},\ and\ \citenamefont
  {Knight}}]{vedral_1997}%
  \BibitemOpen
  \bibfield  {author} {\bibinfo {author} {\bibfnamefont {V.}~\bibnamefont
  {Vedral}}, \bibinfo {author} {\bibfnamefont {M.~B.}\ \bibnamefont {Plenio}},
  \bibinfo {author} {\bibfnamefont {M.~A.}\ \bibnamefont {Rippin}}, \ and\
  \bibinfo {author} {\bibfnamefont {P.~L.}\ \bibnamefont {Knight}},\ }\href
  {\doibase 10.1103/PhysRevLett.78.2275} {\bibfield  {journal} {\bibinfo
  {journal} {Phys. Rev. Lett.}\ }\textbf {\bibinfo {volume} {78}},\ \bibinfo
  {pages} {2275} (\bibinfo {year} {1997})}\BibitemShut {NoStop}%
\bibitem [{\citenamefont {Vedral}\ and\ \citenamefont
  {Plenio}(1998)}]{vedral_1998}%
  \BibitemOpen
  \bibfield  {author} {\bibinfo {author} {\bibfnamefont {V.}~\bibnamefont
  {Vedral}}\ and\ \bibinfo {author} {\bibfnamefont {M.~B.}\ \bibnamefont
  {Plenio}},\ }\href {\doibase 10.1103/PhysRevA.57.1619} {\bibfield  {journal}
  {\bibinfo  {journal} {Phys. Rev. A}\ }\textbf {\bibinfo {volume} {57}},\
  \bibinfo {pages} {1619} (\bibinfo {year} {1998})}\BibitemShut {NoStop}%
\bibitem [{\citenamefont {Vidal}(2000)}]{vidal_2000}%
  \BibitemOpen
  \bibfield  {author} {\bibinfo {author} {\bibfnamefont {G.}~\bibnamefont
  {Vidal}},\ }\href {\doibase 10.1080/09500340008244048} {\bibfield  {journal}
  {\bibinfo  {journal} {J. Mod. Opt.}\ }\textbf {\bibinfo {volume} {47}},\
  \bibinfo {pages} {355} (\bibinfo {year} {2000})}\BibitemShut {NoStop}%
\bibitem [{\citenamefont {Plenio}(2005)}]{plenio_2005}%
  \BibitemOpen
  \bibfield  {author} {\bibinfo {author} {\bibfnamefont {M.~B.}\ \bibnamefont
  {Plenio}},\ }\href {\doibase 10.1103/PhysRevLett.95.090503} {\bibfield
  {journal} {\bibinfo  {journal} {Phys. Rev. Lett.}\ }\textbf {\bibinfo
  {volume} {95}},\ \bibinfo {pages} {090503} (\bibinfo {year}
  {2005})}\BibitemShut {NoStop}%
\bibitem [{\citenamefont {Bengtsson}\ and\ \citenamefont
  {\.{Z}yczkowski}(2007)}]{bengtsson_2007}%
  \BibitemOpen
  \bibfield  {author} {\bibinfo {author} {\bibfnamefont {I.}~\bibnamefont
  {Bengtsson}}\ and\ \bibinfo {author} {\bibfnamefont {K.}~\bibnamefont
  {\.{Z}yczkowski}},\ }\href@noop {} {\emph {\bibinfo {title} {Geometry of
  {{Quantum States}}: {{An Introduction}} to {{Quantum Entanglement}}}}}\
  (\bibinfo  {publisher} {{Cambridge University Press}},\ \bibinfo {year}
  {2007})\BibitemShut {NoStop}%
\bibitem [{\citenamefont {Bennett}\ \emph {et~al.}(1996)\citenamefont
  {Bennett}, \citenamefont {DiVincenzo}, \citenamefont {Smolin},\ and\
  \citenamefont {Wootters}}]{bennett_1996}%
  \BibitemOpen
  \bibfield  {author} {\bibinfo {author} {\bibfnamefont {C.~H.}\ \bibnamefont
  {Bennett}}, \bibinfo {author} {\bibfnamefont {D.~P.}\ \bibnamefont
  {DiVincenzo}}, \bibinfo {author} {\bibfnamefont {J.~A.}\ \bibnamefont
  {Smolin}}, \ and\ \bibinfo {author} {\bibfnamefont {W.~K.}\ \bibnamefont
  {Wootters}},\ }\href {\doibase 10.1103/PhysRevA.54.3824} {\bibfield
  {journal} {\bibinfo  {journal} {Phys. Rev. A}\ }\textbf {\bibinfo {volume}
  {54}},\ \bibinfo {pages} {3824} (\bibinfo {year} {1996})}\BibitemShut
  {NoStop}%
\bibitem [{\citenamefont {Uhlmann}(1998)}]{uhlmann_1998}%
  \BibitemOpen
  \bibfield  {author} {\bibinfo {author} {\bibfnamefont {A.}~\bibnamefont
  {Uhlmann}},\ }\href {\doibase 10.1023/A:1009664331611} {\bibfield  {journal}
  {\bibinfo  {journal} {Open Sys. \& Inf. Dyn.}\ }\textbf {\bibinfo {volume}
  {5}},\ \bibinfo {pages} {209} (\bibinfo {year} {1998})}\BibitemShut {NoStop}%
\bibitem [{\citenamefont {Uhlmann}(2010)}]{uhlmann_2010}%
  \BibitemOpen
  \bibfield  {author} {\bibinfo {author} {\bibfnamefont {A.}~\bibnamefont
  {Uhlmann}},\ }\href {\doibase 10.3390/e12071799} {\bibfield  {journal}
  {\bibinfo  {journal} {Entropy}\ }\textbf {\bibinfo {volume} {12}},\ \bibinfo
  {pages} {1799} (\bibinfo {year} {2010})}\BibitemShut {NoStop}%
\bibitem [{\citenamefont {Wei}\ and\ \citenamefont
  {Goldbart}(2003)}]{wei_2003}%
  \BibitemOpen
  \bibfield  {author} {\bibinfo {author} {\bibfnamefont {T.-C.}\ \bibnamefont
  {Wei}}\ and\ \bibinfo {author} {\bibfnamefont {P.~M.}\ \bibnamefont
  {Goldbart}},\ }\href {\doibase 10.1103/PhysRevA.68.042307} {\bibfield
  {journal} {\bibinfo  {journal} {Phys. Rev. A}\ }\textbf {\bibinfo {volume}
  {68}},\ \bibinfo {pages} {042307} (\bibinfo {year} {2003})}\BibitemShut
  {NoStop}%
\bibitem [{\citenamefont {Streltsov}\ \emph {et~al.}(2010)\citenamefont
  {Streltsov}, \citenamefont {Kampermann},\ and\ \citenamefont
  {Bru{\ss}}}]{streltsov_2010}%
  \BibitemOpen
  \bibfield  {author} {\bibinfo {author} {\bibfnamefont {A.}~\bibnamefont
  {Streltsov}}, \bibinfo {author} {\bibfnamefont {H.}~\bibnamefont
  {Kampermann}}, \ and\ \bibinfo {author} {\bibfnamefont {D.}~\bibnamefont
  {Bru{\ss}}},\ }\href {\doibase 10.1088/1367-2630/12/12/123004} {\bibfield
  {journal} {\bibinfo  {journal} {New J. Phys.}\ }\textbf {\bibinfo {volume}
  {12}},\ \bibinfo {pages} {123004} (\bibinfo {year} {2010})}\BibitemShut
  {NoStop}%
\bibitem [{\citenamefont {D{\"u}r}\ \emph {et~al.}(2000)\citenamefont
  {D{\"u}r}, \citenamefont {Vidal},\ and\ \citenamefont {Cirac}}]{dur_2000}%
  \BibitemOpen
  \bibfield  {author} {\bibinfo {author} {\bibfnamefont {W.}~\bibnamefont
  {D{\"u}r}}, \bibinfo {author} {\bibfnamefont {G.}~\bibnamefont {Vidal}}, \
  and\ \bibinfo {author} {\bibfnamefont {J.~I.}\ \bibnamefont {Cirac}},\ }\href
  {\doibase 10.1103/PhysRevA.62.062314} {\bibfield  {journal} {\bibinfo
  {journal} {Phys. Rev. A}\ }\textbf {\bibinfo {volume} {62}},\ \bibinfo
  {pages} {062314} (\bibinfo {year} {2000})}\BibitemShut {NoStop}%
\bibitem [{\citenamefont {Verstraete}\ \emph {et~al.}(2003)\citenamefont
  {Verstraete}, \citenamefont {Dehaene},\ and\ \citenamefont {{De
  Moor}}}]{verstraete_2003}%
  \BibitemOpen
  \bibfield  {author} {\bibinfo {author} {\bibfnamefont {F.}~\bibnamefont
  {Verstraete}}, \bibinfo {author} {\bibfnamefont {J.}~\bibnamefont {Dehaene}},
  \ and\ \bibinfo {author} {\bibfnamefont {B.}~\bibnamefont {{De Moor}}},\
  }\href {\doibase 10.1103/PhysRevA.68.012103} {\bibfield  {journal} {\bibinfo
  {journal} {Phys. Rev. A}\ }\textbf {\bibinfo {volume} {68}},\ \bibinfo
  {pages} {012103} (\bibinfo {year} {2003})}\BibitemShut {NoStop}%
\bibitem [{\citenamefont {Eltschka}\ \emph {et~al.}(2012)\citenamefont
  {Eltschka}, \citenamefont {Bastin}, \citenamefont {Osterloh},\ and\
  \citenamefont {Siewert}}]{eltschka_2012}%
  \BibitemOpen
  \bibfield  {author} {\bibinfo {author} {\bibfnamefont {C.}~\bibnamefont
  {Eltschka}}, \bibinfo {author} {\bibfnamefont {T.}~\bibnamefont {Bastin}},
  \bibinfo {author} {\bibfnamefont {A.}~\bibnamefont {Osterloh}}, \ and\
  \bibinfo {author} {\bibfnamefont {J.}~\bibnamefont {Siewert}},\ }\href
  {\doibase 10.1103/PhysRevA.85.022301} {\bibfield  {journal} {\bibinfo
  {journal} {Phys. Rev. A}\ }\textbf {\bibinfo {volume} {85}},\ \bibinfo
  {pages} {022301} (\bibinfo {year} {2012})}\BibitemShut {NoStop}%
\bibitem [{\citenamefont {Osterloh}\ and\ \citenamefont
  {Siewert}(2005)}]{osterloh_2005}%
  \BibitemOpen
  \bibfield  {author} {\bibinfo {author} {\bibfnamefont {A.}~\bibnamefont
  {Osterloh}}\ and\ \bibinfo {author} {\bibfnamefont {J.}~\bibnamefont
  {Siewert}},\ }\href {\doibase 10.1103/PhysRevA.72.012337} {\bibfield
  {journal} {\bibinfo  {journal} {Phys. Rev. A}\ }\textbf {\bibinfo {volume}
  {72}},\ \bibinfo {pages} {012337} (\bibinfo {year} {2005})}\BibitemShut
  {NoStop}%
\bibitem [{\citenamefont {Djokovic}\ and\ \citenamefont
  {Osterloh}(2009)}]{djokovic_2009}%
  \BibitemOpen
  \bibfield  {author} {\bibinfo {author} {\bibfnamefont {D.~Z.}\ \bibnamefont
  {Djokovic}}\ and\ \bibinfo {author} {\bibfnamefont {A.}~\bibnamefont
  {Osterloh}},\ }\href {\doibase 10.1063/1.3075830} {\bibfield  {journal}
  {\bibinfo  {journal} {J. Math. Phys.}\ }\textbf {\bibinfo {volume} {50}}
  (\bibinfo {year} {2009}),\ 10.1063/1.3075830}\BibitemShut {NoStop}%
\bibitem [{\citenamefont {Gour}\ and\ \citenamefont
  {Wallach}(2013)}]{gour_2013}%
  \BibitemOpen
  \bibfield  {author} {\bibinfo {author} {\bibfnamefont {G.}~\bibnamefont
  {Gour}}\ and\ \bibinfo {author} {\bibfnamefont {N.~R.}\ \bibnamefont
  {Wallach}},\ }\href {\doibase 10.1103/PhysRevLett.111.060502} {\bibfield
  {journal} {\bibinfo  {journal} {Phys. Rev. Lett.}\ }\textbf {\bibinfo
  {volume} {111}},\ \bibinfo {pages} {060502} (\bibinfo {year}
  {2013})}\BibitemShut {NoStop}%
\bibitem [{\citenamefont {Hill}\ and\ \citenamefont
  {Wootters}(1997)}]{hill_1997}%
  \BibitemOpen
  \bibfield  {author} {\bibinfo {author} {\bibfnamefont {S.}~\bibnamefont
  {Hill}}\ and\ \bibinfo {author} {\bibfnamefont {W.~K.}\ \bibnamefont
  {Wootters}},\ }\href {\doibase 10.1103/PhysRevLett.78.5022} {\bibfield
  {journal} {\bibinfo  {journal} {Phys. Rev. Lett.}\ }\textbf {\bibinfo
  {volume} {78}},\ \bibinfo {pages} {5022} (\bibinfo {year}
  {1997})}\BibitemShut {NoStop}%
\bibitem [{\citenamefont {Wootters}(1998)}]{wootters_1998}%
  \BibitemOpen
  \bibfield  {author} {\bibinfo {author} {\bibfnamefont {W.~K.}\ \bibnamefont
  {Wootters}},\ }\href {\doibase 10.1103/PhysRevLett.80.2245} {\bibfield
  {journal} {\bibinfo  {journal} {Phys. Rev. Lett.}\ }\textbf {\bibinfo
  {volume} {80}},\ \bibinfo {pages} {2245} (\bibinfo {year}
  {1998})}\BibitemShut {NoStop}%
\bibitem [{\citenamefont {Coffman}\ \emph {et~al.}(2000)\citenamefont
  {Coffman}, \citenamefont {Kundu},\ and\ \citenamefont
  {Wootters}}]{coffman_2000}%
  \BibitemOpen
  \bibfield  {author} {\bibinfo {author} {\bibfnamefont {V.}~\bibnamefont
  {Coffman}}, \bibinfo {author} {\bibfnamefont {J.}~\bibnamefont {Kundu}}, \
  and\ \bibinfo {author} {\bibfnamefont {W.~K.}\ \bibnamefont {Wootters}},\
  }\href {\doibase 10.1103/PhysRevA.61.052306} {\bibfield  {journal} {\bibinfo
  {journal} {Phys. Rev. A}\ }\textbf {\bibinfo {volume} {61}},\ \bibinfo
  {pages} {052306} (\bibinfo {year} {2000})}\BibitemShut {NoStop}%
\bibitem [{\citenamefont {Lohmayer}\ \emph {et~al.}(2006)\citenamefont
  {Lohmayer}, \citenamefont {Osterloh}, \citenamefont {Siewert},\ and\
  \citenamefont {Uhlmann}}]{lohmayer_2006}%
  \BibitemOpen
  \bibfield  {author} {\bibinfo {author} {\bibfnamefont {R.}~\bibnamefont
  {Lohmayer}}, \bibinfo {author} {\bibfnamefont {A.}~\bibnamefont {Osterloh}},
  \bibinfo {author} {\bibfnamefont {J.}~\bibnamefont {Siewert}}, \ and\
  \bibinfo {author} {\bibfnamefont {A.}~\bibnamefont {Uhlmann}},\ }\href
  {\doibase 10.1103/PhysRevLett.97.260502} {\bibfield  {journal} {\bibinfo
  {journal} {Phys. Rev. Lett.}\ }\textbf {\bibinfo {volume} {97}},\ \bibinfo
  {pages} {260502} (\bibinfo {year} {2006})}\BibitemShut {NoStop}%
\bibitem [{\citenamefont {Eltschka}\ \emph {et~al.}(2008)\citenamefont
  {Eltschka}, \citenamefont {Osterloh}, \citenamefont {Siewert},\ and\
  \citenamefont {Uhlmann}}]{eltschka_2008}%
  \BibitemOpen
  \bibfield  {author} {\bibinfo {author} {\bibfnamefont {C.}~\bibnamefont
  {Eltschka}}, \bibinfo {author} {\bibfnamefont {A.}~\bibnamefont {Osterloh}},
  \bibinfo {author} {\bibfnamefont {J.}~\bibnamefont {Siewert}}, \ and\
  \bibinfo {author} {\bibfnamefont {A.}~\bibnamefont {Uhlmann}},\ }\href
  {\doibase 10.1088/1367-2630/10/4/043014} {\bibfield  {journal} {\bibinfo
  {journal} {New J. Phys.}\ }\textbf {\bibinfo {volume} {10}},\ \bibinfo
  {pages} {043014} (\bibinfo {year} {2008})}\BibitemShut {NoStop}%
\bibitem [{\citenamefont {Jung}\ \emph
  {et~al.}(2009{\natexlab{a}})\citenamefont {Jung}, \citenamefont {Hwang},
  \citenamefont {Park},\ and\ \citenamefont {Son}}]{jung_2009}%
  \BibitemOpen
  \bibfield  {author} {\bibinfo {author} {\bibfnamefont {E.}~\bibnamefont
  {Jung}}, \bibinfo {author} {\bibfnamefont {M.-R.}\ \bibnamefont {Hwang}},
  \bibinfo {author} {\bibfnamefont {D.~K.}\ \bibnamefont {Park}}, \ and\
  \bibinfo {author} {\bibfnamefont {J.-W.}\ \bibnamefont {Son}},\ }\href
  {\doibase 10.1103/PhysRevA.79.024306} {\bibfield  {journal} {\bibinfo
  {journal} {Phys. Rev. A}\ }\textbf {\bibinfo {volume} {79}},\ \bibinfo
  {pages} {024306} (\bibinfo {year} {2009}{\natexlab{a}})}\BibitemShut
  {NoStop}%
\bibitem [{\citenamefont {Jung}\ \emph
  {et~al.}(2009{\natexlab{b}})\citenamefont {Jung}, \citenamefont {Park},\ and\
  \citenamefont {Son}}]{jung_2009-1}%
  \BibitemOpen
  \bibfield  {author} {\bibinfo {author} {\bibfnamefont {E.}~\bibnamefont
  {Jung}}, \bibinfo {author} {\bibfnamefont {D.~K.}\ \bibnamefont {Park}}, \
  and\ \bibinfo {author} {\bibfnamefont {J.-W.}\ \bibnamefont {Son}},\ }\href
  {\doibase 10.1103/PhysRevA.80.010301} {\bibfield  {journal} {\bibinfo
  {journal} {Phys. Rev. A}\ }\textbf {\bibinfo {volume} {80}},\ \bibinfo
  {pages} {010301} (\bibinfo {year} {2009}{\natexlab{b}})}\BibitemShut
  {NoStop}%
\bibitem [{\citenamefont {He}\ \emph {et~al.}(2011)\citenamefont {He},
  \citenamefont {Wang}, \citenamefont {Fei}, \citenamefont {Sun},\ and\
  \citenamefont {Wen}}]{he_2011}%
  \BibitemOpen
  \bibfield  {author} {\bibinfo {author} {\bibfnamefont {S.-J.}\ \bibnamefont
  {He}}, \bibinfo {author} {\bibfnamefont {X.-H.}\ \bibnamefont {Wang}},
  \bibinfo {author} {\bibfnamefont {S.-M.}\ \bibnamefont {Fei}}, \bibinfo
  {author} {\bibfnamefont {H.-X.}\ \bibnamefont {Sun}}, \ and\ \bibinfo
  {author} {\bibfnamefont {Q.-Y.}\ \bibnamefont {Wen}},\ }\href {\doibase
  10.1088/0253-6102/55/2/12} {\bibfield  {journal} {\bibinfo  {journal}
  {Commun. Theor. Phys.}\ }\textbf {\bibinfo {volume} {55}},\ \bibinfo {pages}
  {251} (\bibinfo {year} {2011})}\BibitemShut {NoStop}%
\bibitem [{\citenamefont {Siewert}\ and\ \citenamefont
  {Eltschka}(2012)}]{siewert_2012}%
  \BibitemOpen
  \bibfield  {author} {\bibinfo {author} {\bibfnamefont {J.}~\bibnamefont
  {Siewert}}\ and\ \bibinfo {author} {\bibfnamefont {C.}~\bibnamefont
  {Eltschka}},\ }\href {\doibase 10.1103/PhysRevLett.108.230502} {\bibfield
  {journal} {\bibinfo  {journal} {Phys. Rev. Lett.}\ }\textbf {\bibinfo
  {volume} {108}},\ \bibinfo {pages} {230502} (\bibinfo {year}
  {2012})}\BibitemShut {NoStop}%
\bibitem [{\citenamefont {Viehmann}\ \emph {et~al.}(2012)\citenamefont
  {Viehmann}, \citenamefont {Eltschka},\ and\ \citenamefont
  {Siewert}}]{viehmann_2012}%
  \BibitemOpen
  \bibfield  {author} {\bibinfo {author} {\bibfnamefont {O.}~\bibnamefont
  {Viehmann}}, \bibinfo {author} {\bibfnamefont {C.}~\bibnamefont {Eltschka}},
  \ and\ \bibinfo {author} {\bibfnamefont {J.}~\bibnamefont {Siewert}},\ }\href
  {\doibase 10.1007/s00340-011-4864-x} {\bibfield  {journal} {\bibinfo
  {journal} {Appl. Phys. B}\ }\textbf {\bibinfo {volume} {106}},\ \bibinfo
  {pages} {533} (\bibinfo {year} {2012})}\BibitemShut {NoStop}%
\bibitem [{\citenamefont {Eltschka}\ and\ \citenamefont
  {Siewert}(2012)}]{eltschka_2012-1}%
  \BibitemOpen
  \bibfield  {author} {\bibinfo {author} {\bibfnamefont {C.}~\bibnamefont
  {Eltschka}}\ and\ \bibinfo {author} {\bibfnamefont {J.}~\bibnamefont
  {Siewert}},\ }\href@noop {} {\bibfield  {journal} {\bibinfo  {journal} {Sci.
  Rep.}\ }\textbf {\bibinfo {volume} {2}} (\bibinfo {year} {2012})}\BibitemShut
  {NoStop}%
\bibitem [{\citenamefont {Regula}\ and\ \citenamefont
  {Adesso}(2016)}]{regula_2016}%
  \BibitemOpen
  \bibfield  {author} {\bibinfo {author} {\bibfnamefont {B.}~\bibnamefont
  {Regula}}\ and\ \bibinfo {author} {\bibfnamefont {G.}~\bibnamefont
  {Adesso}},\ }\href {\doibase 10.1103/PhysRevLett.116.070504} {\bibfield
  {journal} {\bibinfo  {journal} {Phys. Rev. Lett.}\ }\textbf {\bibinfo
  {volume} {116}},\ \bibinfo {pages} {070504} (\bibinfo {year}
  {2016})}\BibitemShut {NoStop}%
\bibitem [{\citenamefont {Jung}\ and\ \citenamefont {Park}(2015)}]{jung_2015}%
  \BibitemOpen
  \bibfield  {author} {\bibinfo {author} {\bibfnamefont {E.}~\bibnamefont
  {Jung}}\ and\ \bibinfo {author} {\bibfnamefont {D.}~\bibnamefont {Park}},\
  }\href {\doibase 10.1007/s11128-015-1039-4} {\bibfield  {journal} {\bibinfo
  {journal} {Quant. Inf. Proc.}\ }\textbf {\bibinfo {volume} {14}},\ \bibinfo
  {pages} {3317} (\bibinfo {year} {2015})}\BibitemShut {NoStop}%
\bibitem [{\citenamefont {Osterloh}\ \emph {et~al.}(2008)\citenamefont
  {Osterloh}, \citenamefont {Siewert},\ and\ \citenamefont
  {Uhlmann}}]{osterloh_2008}%
  \BibitemOpen
  \bibfield  {author} {\bibinfo {author} {\bibfnamefont {A.}~\bibnamefont
  {Osterloh}}, \bibinfo {author} {\bibfnamefont {J.}~\bibnamefont {Siewert}}, \
  and\ \bibinfo {author} {\bibfnamefont {A.}~\bibnamefont {Uhlmann}},\ }\href
  {\doibase 10.1103/PhysRevA.77.032310} {\bibfield  {journal} {\bibinfo
  {journal} {Phys. Rev. A}\ }\textbf {\bibinfo {volume} {77}},\ \bibinfo
  {pages} {032310} (\bibinfo {year} {2008})}\BibitemShut {NoStop}%
\bibitem [{\citenamefont {Carath\'{e}odory}(1954)}]{caratheodory_1954}%
  \BibitemOpen
  \bibfield  {author} {\bibinfo {author} {\bibfnamefont {C.}~\bibnamefont
  {Carath\'{e}odory}},\ }\href@noop {} {\emph {\bibinfo {title} {Theory of
  functions of a complex variable}}},\ Vol.~\bibinfo {volume} {1}\ (\bibinfo
  {publisher} {Chelsea Publishing Co.},\ \bibinfo {year} {1954})\BibitemShut
  {NoStop}%
\bibitem [{\citenamefont {Whyte}(1952)}]{whyte_1952}%
  \BibitemOpen
  \bibfield  {author} {\bibinfo {author} {\bibfnamefont {L.~L.}\ \bibnamefont
  {Whyte}},\ }\href@noop {} {\bibfield  {journal} {\bibinfo  {journal} {Amer.
  Math. Monthly}\ }\textbf {\bibinfo {volume} {59}},\ \bibinfo {pages} {606}
  (\bibinfo {year} {1952})}\BibitemShut {NoStop}%
\bibitem [{\citenamefont {Wagner}(1989)}]{wagner_1989}%
  \BibitemOpen
  \bibfield  {author} {\bibinfo {author} {\bibfnamefont {G.}~\bibnamefont
  {Wagner}},\ }\href {\doibase 10.1017/S1446788700033206} {\bibfield  {journal}
  {\bibinfo  {journal} {J. Austral. Math. Soc. (Series A)}\ }\textbf {\bibinfo
  {volume} {47}},\ \bibinfo {pages} {466} (\bibinfo {year} {1989})}\BibitemShut
  {NoStop}%
\bibitem [{\citenamefont {Rakhmanov}\ \emph {et~al.}(1994)\citenamefont
  {Rakhmanov}, \citenamefont {Saff},\ and\ \citenamefont
  {Zhou}}]{rakhmanov_1994}%
  \BibitemOpen
  \bibfield  {author} {\bibinfo {author} {\bibfnamefont {E.~A.}\ \bibnamefont
  {Rakhmanov}}, \bibinfo {author} {\bibfnamefont {E.~B.}\ \bibnamefont {Saff}},
  \ and\ \bibinfo {author} {\bibfnamefont {Y.~M.}\ \bibnamefont {Zhou}},\
  }\href@noop {} {\bibfield  {journal} {\bibinfo  {journal} {Math. Res. Lett.}\
  }\textbf {\bibinfo {volume} {1}},\ \bibinfo {pages} {647–662} (\bibinfo
  {year} {1994})}\BibitemShut {NoStop}%
\bibitem [{\citenamefont {Wootters}(2001)}]{wootters_2001}%
  \BibitemOpen
  \bibfield  {author} {\bibinfo {author} {\bibfnamefont {W.~K.}\ \bibnamefont
  {Wootters}},\ }\href@noop {} {\bibfield  {journal} {\bibinfo  {journal}
  {Quant. Inf. Comp.}\ }\textbf {\bibinfo {volume} {1}},\ \bibinfo {pages} {27}
  (\bibinfo {year} {2001})}\BibitemShut {NoStop}%
\bibitem [{\citenamefont {Cassini}(1684)}]{cassini_1684}%
  \BibitemOpen
  \bibfield  {author} {\bibinfo {author} {\bibfnamefont {G.~D.}\ \bibnamefont
  {Cassini}},\ }\href@noop {} {\emph {\bibinfo {title} {Les \'{e}l\'{e}mens de
  l'astronomie}}}\ (\bibinfo  {publisher} {L'Imprimerie Royale},\ \bibinfo
  {address} {Paris},\ \bibinfo {year} {1684})\BibitemShut {NoStop}%
\bibitem [{\citenamefont {Needham}(1998)}]{needham_1998}%
  \BibitemOpen
  \bibfield  {author} {\bibinfo {author} {\bibfnamefont {T.}~\bibnamefont
  {Needham}},\ }\href@noop {} {\emph {\bibinfo {title} {Visual Complex
  Analysis}}}\ (\bibinfo  {publisher} {Clarendon Press},\ \bibinfo {year}
  {1998})\BibitemShut {NoStop}%
\bibitem [{\citenamefont {Rockafellar}(1970)}]{rockafellar_1970}%
  \BibitemOpen
  \bibfield  {author} {\bibinfo {author} {\bibfnamefont {R.}~\bibnamefont
  {Rockafellar}},\ }\href@noop {} {\emph {\bibinfo {title} {Convex Analysis}}}\
  (\bibinfo  {publisher} {Princeton University Press},\ \bibinfo {year}
  {1970})\BibitemShut {NoStop}%
\bibitem [{\citenamefont {Lewenstein}\ and\ \citenamefont
  {Sanpera}(1998)}]{lewenstein_1998}%
  \BibitemOpen
  \bibfield  {author} {\bibinfo {author} {\bibfnamefont {M.}~\bibnamefont
  {Lewenstein}}\ and\ \bibinfo {author} {\bibfnamefont {A.}~\bibnamefont
  {Sanpera}},\ }\href {\doibase 10.1103/PhysRevLett.80.2261} {\bibfield
  {journal} {\bibinfo  {journal} {Phys. Rev. Lett.}\ }\textbf {\bibinfo
  {volume} {80}},\ \bibinfo {pages} {2261} (\bibinfo {year}
  {1998})}\BibitemShut {NoStop}%
\bibitem [{\citenamefont {Rodriques}\ \emph {et~al.}(2014)\citenamefont
  {Rodriques}, \citenamefont {Datta},\ and\ \citenamefont
  {Love}}]{rodriques_2014}%
  \BibitemOpen
  \bibfield  {author} {\bibinfo {author} {\bibfnamefont {S.}~\bibnamefont
  {Rodriques}}, \bibinfo {author} {\bibfnamefont {N.}~\bibnamefont {Datta}}, \
  and\ \bibinfo {author} {\bibfnamefont {P.}~\bibnamefont {Love}},\ }\href
  {\doibase 10.1103/PhysRevA.90.012340} {\bibfield  {journal} {\bibinfo
  {journal} {Phys. Rev. A}\ }\textbf {\bibinfo {volume} {90}},\ \bibinfo
  {pages} {012340} (\bibinfo {year} {2014})}\BibitemShut {NoStop}%
\bibitem [{\citenamefont {Osterloh}(2016)}]{osterloh_2016}%
  \BibitemOpen
  \bibfield  {author} {\bibinfo {author} {\bibfnamefont {A.}~\bibnamefont
  {Osterloh}},\ }\href {\doibase 10.1103/PhysRevA.93.052322} {\bibfield
  {journal} {\bibinfo  {journal} {Phys. Rev. A}\ }\textbf {\bibinfo {volume}
  {93}},\ \bibinfo {pages} {052322} (\bibinfo {year} {2016})}\BibitemShut
  {NoStop}%
\bibitem [{\citenamefont {Baumgratz}\ \emph {et~al.}(2014)\citenamefont
  {Baumgratz}, \citenamefont {Cramer},\ and\ \citenamefont
  {Plenio}}]{baumgratz_2014}%
  \BibitemOpen
  \bibfield  {author} {\bibinfo {author} {\bibfnamefont {T.}~\bibnamefont
  {Baumgratz}}, \bibinfo {author} {\bibfnamefont {M.}~\bibnamefont {Cramer}}, \
  and\ \bibinfo {author} {\bibfnamefont {M.}~\bibnamefont {Plenio}},\ }\href
  {\doibase 10.1103/PhysRevLett.113.140401} {\bibfield  {journal} {\bibinfo
  {journal} {Phys. Rev. Lett.}\ }\textbf {\bibinfo {volume} {113}},\ \bibinfo
  {pages} {140401} (\bibinfo {year} {2014})}\BibitemShut {NoStop}%
\bibitem [{\citenamefont {Eltschka}\ and\ \citenamefont
  {Siewert}(2014{\natexlab{b}})}]{eltschka_2014}%
  \BibitemOpen
  \bibfield  {author} {\bibinfo {author} {\bibfnamefont {C.}~\bibnamefont
  {Eltschka}}\ and\ \bibinfo {author} {\bibfnamefont {J.}~\bibnamefont
  {Siewert}},\ }\href {\doibase 10.1103/PhysRevA.89.022312} {\bibfield
  {journal} {\bibinfo  {journal} {Phys. Rev. A}\ }\textbf {\bibinfo {volume}
  {89}},\ \bibinfo {pages} {022312} (\bibinfo {year}
  {2014}{\natexlab{b}})}\BibitemShut {NoStop}%
\bibitem [{\citenamefont {Ac{\'\i}n}\ \emph {et~al.}(2001)\citenamefont
  {Ac{\'\i}n}, \citenamefont {Bru{\ss}}, \citenamefont {Lewenstein},\ and\
  \citenamefont {Sanpera}}]{acin_2001}%
  \BibitemOpen
  \bibfield  {author} {\bibinfo {author} {\bibfnamefont {A.}~\bibnamefont
  {Ac{\'\i}n}}, \bibinfo {author} {\bibfnamefont {D.}~\bibnamefont {Bru{\ss}}},
  \bibinfo {author} {\bibfnamefont {M.}~\bibnamefont {Lewenstein}}, \ and\
  \bibinfo {author} {\bibfnamefont {A.}~\bibnamefont {Sanpera}},\ }\href
  {\doibase 10.1103/PhysRevLett.87.040401} {\bibfield  {journal} {\bibinfo
  {journal} {Phys. Rev. Lett.}\ }\textbf {\bibinfo {volume} {87}},\ \bibinfo
  {pages} {040401} (\bibinfo {year} {2001})}\BibitemShut {NoStop}%
\bibitem [{\citenamefont {Verstraete}\ \emph {et~al.}(2002)\citenamefont
  {Verstraete}, \citenamefont {Dehaene}, \citenamefont {{De Moor}},\ and\
  \citenamefont {Verschelde}}]{verstraete_2002}%
  \BibitemOpen
  \bibfield  {author} {\bibinfo {author} {\bibfnamefont {F.}~\bibnamefont
  {Verstraete}}, \bibinfo {author} {\bibfnamefont {J.}~\bibnamefont {Dehaene}},
  \bibinfo {author} {\bibfnamefont {B.}~\bibnamefont {{De Moor}}}, \ and\
  \bibinfo {author} {\bibfnamefont {H.}~\bibnamefont {Verschelde}},\ }\href
  {\doibase 10.1103/PhysRevA.65.052112} {\bibfield  {journal} {\bibinfo
  {journal} {Phys. Rev. A}\ }\textbf {\bibinfo {volume} {65}},\ \bibinfo
  {pages} {052112} (\bibinfo {year} {2002})}\BibitemShut {NoStop}%
\bibitem [{\citenamefont {Chterental}\ and\ \citenamefont
  {Djokovic}(2007)}]{chterental_2007}%
  \BibitemOpen
  \bibfield  {author} {\bibinfo {author} {\bibfnamefont {O.}~\bibnamefont
  {Chterental}}\ and\ \bibinfo {author} {\bibfnamefont {D.~Z.}\ \bibnamefont
  {Djokovic}},\ }in\ \href@noop {} {\emph {\bibinfo {booktitle} {Linear
  {{Algebra Research Advances}}}}},\ \bibinfo {editor} {edited by\ \bibinfo
  {editor} {\bibfnamefont {G.~D.}\ \bibnamefont {Ling}}}\ (\bibinfo
  {publisher} {{Nova Science Publishers}},\ \bibinfo {address} {New York},\
  \bibinfo {year} {2007})\ pp.\ \bibinfo {pages} {133--167}\BibitemShut
  {NoStop}%
\bibitem [{\citenamefont {Spee}\ \emph {et~al.}(2016)\citenamefont {Spee},
  \citenamefont {de~Vicente},\ and\ \citenamefont {Kraus}}]{spee_2015}%
  \BibitemOpen
  \bibfield  {author} {\bibinfo {author} {\bibfnamefont {C.}~\bibnamefont
  {Spee}}, \bibinfo {author} {\bibfnamefont {J.~I.}\ \bibnamefont
  {de~Vicente}}, \ and\ \bibinfo {author} {\bibfnamefont {B.}~\bibnamefont
  {Kraus}},\ }\href {\doibase 10.1063/1.4946895} {\bibfield  {journal}
  {\bibinfo  {journal} {J. Math. Phys.}\ }\textbf {\bibinfo {volume} {57}},\
  \bibinfo {pages} {052201} (\bibinfo {year} {2016})}\BibitemShut {NoStop}%
\bibitem [{\citenamefont {Holweck}\ \emph {et~al.}(2014)\citenamefont
  {Holweck}, \citenamefont {Luque},\ and\ \citenamefont
  {Thibon}}]{holweck_2014}%
  \BibitemOpen
  \bibfield  {author} {\bibinfo {author} {\bibfnamefont {F.}~\bibnamefont
  {Holweck}}, \bibinfo {author} {\bibfnamefont {J.-G.}\ \bibnamefont {Luque}},
  \ and\ \bibinfo {author} {\bibfnamefont {J.-Y.}\ \bibnamefont {Thibon}},\
  }\href {\doibase 10.1063/1.4858336} {\bibfield  {journal} {\bibinfo
  {journal} {J. Math. Phys.}\ }\textbf {\bibinfo {volume} {55}},\ \bibinfo
  {eid} {012202} (\bibinfo {year} {2014}),\ 10.1063/1.4858336}\BibitemShut
  {NoStop}%
\bibitem [{\citenamefont {Holweck}\ \emph {et~al.}(2016)\citenamefont
  {Holweck}, \citenamefont {Luque},\ and\ \citenamefont
  {Thibon}}]{holweck_2016}%
  \BibitemOpen
  \bibfield  {author} {\bibinfo {author} {\bibfnamefont {F.}~\bibnamefont
  {Holweck}}, \bibinfo {author} {\bibfnamefont {J.-G.}\ \bibnamefont {Luque}},
  \ and\ \bibinfo {author} {\bibfnamefont {J.-Y.}\ \bibnamefont {Thibon}},\
  }\href@noop {} {} (\bibinfo {year} {2016}),\ \Eprint
  {http://arxiv.org/abs/1606.05569} {arXiv:1606.05569 [math-ph]} \BibitemShut
  {NoStop}%
\bibitem [{\citenamefont {Brylinski}\ and\ \citenamefont
  {Brylinski}(2002)}]{brylinski_2002}%
  \BibitemOpen
  \bibfield  {author} {\bibinfo {author} {\bibfnamefont {J.~L.}\ \bibnamefont
  {Brylinski}}\ and\ \bibinfo {author} {\bibfnamefont {R.}~\bibnamefont
  {Brylinski}},\ }in\ \href@noop {} {\emph {\bibinfo {booktitle} {Mathematics
  of Quantum Computation}}}\ (\bibinfo  {publisher} {Chapman \& Hall/CRC
  Press},\ \bibinfo {year} {2002})\BibitemShut {NoStop}%
\bibitem [{\citenamefont {Terhal}(2004)}]{terhal_2004}%
  \BibitemOpen
  \bibfield  {author} {\bibinfo {author} {\bibfnamefont {B.}~\bibnamefont
  {Terhal}},\ }\href {\doibase 10.1147/rd.481.0071} {\bibfield  {journal}
  {\bibinfo  {journal} {IBM J. Res. Dev.}\ }\textbf {\bibinfo {volume} {48}},\
  \bibinfo {pages} {71} (\bibinfo {year} {2004})}\BibitemShut {NoStop}%
\bibitem [{\citenamefont {Streltsov}\ \emph {et~al.}(2012)\citenamefont
  {Streltsov}, \citenamefont {Adesso}, \citenamefont {Piani},\ and\
  \citenamefont {Bru{\ss}}}]{streltsov_2012}%
  \BibitemOpen
  \bibfield  {author} {\bibinfo {author} {\bibfnamefont {A.}~\bibnamefont
  {Streltsov}}, \bibinfo {author} {\bibfnamefont {G.}~\bibnamefont {Adesso}},
  \bibinfo {author} {\bibfnamefont {M.}~\bibnamefont {Piani}}, \ and\ \bibinfo
  {author} {\bibfnamefont {D.}~\bibnamefont {Bru{\ss}}},\ }\href {\doibase
  10.1103/PhysRevLett.109.050503} {\bibfield  {journal} {\bibinfo  {journal}
  {Phys. Rev. Lett.}\ }\textbf {\bibinfo {volume} {109}},\ \bibinfo {pages}
  {050503} (\bibinfo {year} {2012})}\BibitemShut {NoStop}%
\bibitem [{\citenamefont {Eltschka}\ and\ \citenamefont
  {Siewert}(2015)}]{eltschka_2015}%
  \BibitemOpen
  \bibfield  {author} {\bibinfo {author} {\bibfnamefont {C.}~\bibnamefont
  {Eltschka}}\ and\ \bibinfo {author} {\bibfnamefont {J.}~\bibnamefont
  {Siewert}},\ }\href {\doibase 10.1103/PhysRevLett.114.140402} {\bibfield
  {journal} {\bibinfo  {journal} {Phys. Rev. Lett.}\ }\textbf {\bibinfo
  {volume} {114}},\ \bibinfo {pages} {140402} (\bibinfo {year}
  {2015})}\BibitemShut {NoStop}%
\bibitem [{\citenamefont {Lancien}\ \emph {et~al.}(2016)\citenamefont
  {Lancien}, \citenamefont {{Di Martino}}, \citenamefont {Huber}, \citenamefont
  {Piani}, \citenamefont {Adesso},\ and\ \citenamefont
  {Winter}}]{lancien_2016}%
  \BibitemOpen
  \bibfield  {author} {\bibinfo {author} {\bibfnamefont {C.}~\bibnamefont
  {Lancien}}, \bibinfo {author} {\bibfnamefont {S.}~\bibnamefont {{Di
  Martino}}}, \bibinfo {author} {\bibfnamefont {M.}~\bibnamefont {Huber}},
  \bibinfo {author} {\bibfnamefont {M.}~\bibnamefont {Piani}}, \bibinfo
  {author} {\bibfnamefont {G.}~\bibnamefont {Adesso}}, \ and\ \bibinfo {author}
  {\bibfnamefont {A.}~\bibnamefont {Winter}},\ }\href {\doibase
  10.1103/PhysRevLett.117.060501} {\bibfield  {journal} {\bibinfo  {journal}
  {Phys. Rev. Lett.}\ }\textbf {\bibinfo {volume} {117}},\ \bibinfo {pages}
  {060501} (\bibinfo {year} {2016})}\BibitemShut {NoStop}%
\bibitem [{\citenamefont {Osborne}\ and\ \citenamefont
  {Verstraete}(2006)}]{osborne_2006}%
  \BibitemOpen
  \bibfield  {author} {\bibinfo {author} {\bibfnamefont {T.~J.}\ \bibnamefont
  {Osborne}}\ and\ \bibinfo {author} {\bibfnamefont {F.}~\bibnamefont
  {Verstraete}},\ }\href {\doibase 10.1103/PhysRevLett.96.220503} {\bibfield
  {journal} {\bibinfo  {journal} {Phys. Rev. Lett.}\ }\textbf {\bibinfo
  {volume} {96}},\ \bibinfo {pages} {220503} (\bibinfo {year}
  {2006})}\BibitemShut {NoStop}%
\bibitem [{\citenamefont {Regula}\ \emph {et~al.}(2014)\citenamefont {Regula},
  \citenamefont {{Di Martino}}, \citenamefont {Lee},\ and\ \citenamefont
  {Adesso}}]{regula_2014}%
  \BibitemOpen
  \bibfield  {author} {\bibinfo {author} {\bibfnamefont {B.}~\bibnamefont
  {Regula}}, \bibinfo {author} {\bibfnamefont {S.}~\bibnamefont {{Di
  Martino}}}, \bibinfo {author} {\bibfnamefont {S.}~\bibnamefont {Lee}}, \ and\
  \bibinfo {author} {\bibfnamefont {G.}~\bibnamefont {Adesso}},\ }\href
  {\doibase 10.1103/PhysRevLett.113.110501} {\bibfield  {journal} {\bibinfo
  {journal} {Phys. Rev. Lett.}\ }\textbf {\bibinfo {volume} {113}},\ \bibinfo
  {pages} {110501} (\bibinfo {year} {2014})}\BibitemShut {NoStop}%
\bibitem [{\citenamefont {Regula}\ \emph {et~al.}(2016)\citenamefont {Regula},
  \citenamefont {Osterloh},\ and\ \citenamefont {Adesso}}]{regula_2016-1}%
  \BibitemOpen
  \bibfield  {author} {\bibinfo {author} {\bibfnamefont {B.}~\bibnamefont
  {Regula}}, \bibinfo {author} {\bibfnamefont {A.}~\bibnamefont {Osterloh}}, \
  and\ \bibinfo {author} {\bibfnamefont {G.}~\bibnamefont {Adesso}},\ }\href
  {\doibase 10.1103/PhysRevA.93.052338} {\bibfield  {journal} {\bibinfo
  {journal} {Phys. Rev. A}\ }\textbf {\bibinfo {volume} {93}},\ \bibinfo
  {pages} {052338} (\bibinfo {year} {2016})}\BibitemShut {NoStop}%
\bibitem [{\citenamefont {Kimura}(2003)}]{kimura_2003}%
  \BibitemOpen
  \bibfield  {author} {\bibinfo {author} {\bibfnamefont {G.}~\bibnamefont
  {Kimura}},\ }\href {\doibase 10.1016/S0375-9601(03)00941-1} {\bibfield
  {journal} {\bibinfo  {journal} {Phys. Lett. A}\ }\textbf {\bibinfo {volume}
  {314}},\ \bibinfo {pages} {339} (\bibinfo {year} {2003})}\BibitemShut
  {NoStop}%
\end{thebibliography}%

\end{document}